\documentclass[12pt]{article}
\usepackage{amssymb,amsmath,epsfig}
\usepackage{graphicx}
\usepackage{amsfonts,graphicx,amssymb,amsmath,amsthm,epsfig}
\usepackage{float}
\allowdisplaybreaks
\begin{document}
\title{\textbf{Impact of Pulsar SAX J1748.9-2021 Observations on
$f(\mathcal{Q}, \mathbb{T})$ Gravity}}
\author{M. Sharif$^{1,2}$\thanks{msharif.math@pu.edu.pk}~ and
Iqra Ibrar$^1$\thanks{iqraibrar26@gmail.com}\\
$^1$Department of Mathematics and Statistics, The University of Lahore,\\
1-KM Defence Road Lahore-54000, Pakistan.\\
$^2$ Research Center of Astrophysics and Cosmology, Khazar
University,\\ Baku, AZ1096, 41 Mehseti Street, Azerbaijan.}
\date{}
\maketitle

\begin{abstract}
The main objective of this study is to investigate the viability and
stability of a pulsar filled with anisotropic matter in
$f(\mathcal{Q}, \mathbb{T})$ gravity, where $\mathcal{Q}$ represents
non-metricity and $\mathbb{T}$ is the trace of the energy-momentum
tensor. In this context, we employ non-singular solution and a
particular model of this gravity. We use junction conditions to
evaluate unknown constants in the metric coefficients. Observations
from the pulsar SAX J1748.9-2021 star are employed to validate the
model, producing stable configurations that address both geometric
and physical characteristics. This framework establishes
relationships between various physical quantities, including fluid
parameters, anisotropy, mass-radius relation, redshift, the
Zeldovich condition, energy conditions, causality conditions,
adiabatic index, Tolman-Oppenheimer-Volkoff equation, the equation
of state parameter, and compactness. Our findings confirm the
viability and stability of the proposed pulsar star in this
theoretical framework.
\end{abstract}

\section{Introduction}

General relativity (GR), proposed by Albert Einstein, is one of the
most groundbreaking scientific achievements of the 20th century and
serves as a cornerstone of modern physics. It provides a
comprehensive framework to understand the evolution of the universe
and its hidden mysteries. Recent advancements in cosmology have
introduced innovative methods to study the universe accelerated
expansion, supported by observational evidence such as redshift
supernova observations \cite{1-b}, large-scale structures \cite{1-c}
and variations in cosmic microwave background radiation \cite{1-d}.
This rapid expansion is attributed to dark energy (DE), an unknown
force responsible for approximately 68\% of the universe total
energy. While GR provides significant insights into this phenomenon
through the inclusion of cosmological constant, it also presents
challenges like fine-tuning and the coincidence problem. To address
these limitations, modified gravity theories have been proposed,
which expand on GR and offer explanation for late-time cosmic
evolution and the accelerated expansion \cite{1-e}. These theories
align with GR in weak-field conditions but show notable deviations
in strong gravitational fields. Compact astrophysical objects such
as neutron stars, which exist in these extreme gravitational
regimes, provide ideal environments to test the predictions of these
modified theories and uncover phenomena unexplained by GR. Thus,
while GR remains a fundamental pillar validated by extensive
observations, modified theories open promising pathways for
advancing our understanding of stellar astrophysics and cosmology.

One way to extend GR is by adopting broader geometric frameworks
rather than Riemannian metric space. These new frameworks have the
potential to elucidate the gravitational field and accurately
describe the behavior of matter on vast cosmic scales. Weyl
\cite{3-a} developed a more general geometry than Riemannian space,
aiming to unify fundamental forces within a single geometric
structure. A fundamental concept in Riemannian geometry is the
Levi-Civita connection, which enables the comparison of vectors by
their length. Weyl implemented a connection for parallel transport
that disregards information about the length of vectors. He added
another connection called the length connection, which adjusts the
conformal factor instead of the direction of vector transport. The
theory claims that the covariant divergence of the metric tensor is
non-zero, introducing the concept of non-metricity. He correlated
the length connection with the electromagnetic potential and applied
it to physical context.

The torsion and non-metricity scalars are mathematical quantities
that emerge in modified theories incorporating non-Riemannian
geometry. The $f(\mathrm{T})$ gravity, a modified version of GR uses
torsion to describe gravitational interactions. Nester and Yo
\cite{11c} investigated teleparallel geometry characterized by zero
curvature and torsion, but with a non-metricity included. This
framework is known as symmetric teleparallel equivalent of GR (STG)
and its extension is referred to as $f(\mathcal{Q})$ gravity
\cite{12}. Lazkoz et al. \cite{5-d} investigated the constraints of
$f(\mathcal{Q})$ theory using a polynomial redshift function. The
energy conditions for two distinct $f(\mathcal{Q})$ gravity models
have been analyzed in \cite{5-e}. Shekh \cite{a} examined the
dynamics of a DE model within this framework. Lin and Zhai \cite{b}
investigated the impact of the theory on the internal and external
structures of compact stars by analyzing spherically symmetric
solutions. Frusciante \cite{c} proposed a model in $f(\mathcal{Q})$
gravity that aligns with $\Lambda$ cold dark matter (CDM) on a large
scale while exhibiting distinct variations at smaller scales,
offering measurable predictions. Dimakis et al. \cite{d} studied the
universe evolution both with and without a cosmological constant, as
well as exploring the implications of phantom DE. Sokoliuk et al.
\cite{e} utilized the Pantheon dataset to explore cosmic evolution
under $f(\mathcal{Q})$ gravity. Bhar and Pretel \cite{f} focused on
compact stars and DE stars using the singularity-free
Tolman-Kuchowicz metric, investigating quark stars by calculating
their mass and radius based on specific equations of state (EoS).
The pulsar SAX J1808.4-3658 was instrumental in analyzing the
physical properties of DE stars, including their stability and
causality conditions. Recent studies \cite{g} have extensively
explored various geometric and physical aspects of $f(\mathcal{Q})$
gravity, providing valuable insights into the theory's versatility.
Sharif and Ajmal \cite{5-g} explored the characteristics of
generalized ghost DE in $f(\mathcal{Q})$ theory, while Sharif et al.
\cite{5-h} explored the concept of a cosmological bounce.

Xu et al. \cite{6-a} extended the $f(\mathcal{Q})$ theory by
incorporating the trace of the energy-momentum tensor (EMT) into the
functional action, resulting in the $f(\mathcal{Q}, \mathbb{T})$
gravity. This theory establishes a specific connection between the
trace of the EMT and non-metricity. This coupling offers a unified
approach to gravitational phenomena, enabling $f(\mathcal{Q},
\mathbb{T})$ gravity to explain both early-time inflation and
late-time cosmic acceleration without requiring exotic DE. By
modifying spacetime geometry itself, the theory naturally accounts
for cosmic acceleration, incorporating matter density and pressure
effects through $\mathbb{T}$. Unlike other modified theories of
gravity that rely on additional fields or excessive fine-tuning,
$f(\mathcal{Q}, \mathbb{T})$ gravity enhances the flexibility of
cosmological modeling. It fits observational data effectively while
predicting large-scale cosmic behavior. Additionally, its
second-order field equations simplify calculations, making it
practical for theoretical and computational studies. These
advantages position $f(\mathcal{Q}, \mathbb{T})$ gravity as a
promising alternative framework for advancing cosmological
understanding, though further research is necessary to establish its
validity as a comprehensive model of the universe dynamics.

The motivation for studying this theory includes exploring its theoretical
implications, checking its consistency with observational data and understanding its
importance in cosmological applications. Arora et al. \cite{6-b} examined the
potential of this theory to explain the cosmic late-time acceleration without the
necessity of additional DE forms. Bhattacharjee \cite{gg} investigated gravitational
baryogenesis in this gravity, finding that it can significantly enhance baryon to
entropy ratio. Godani and Samanta \cite{hh} explored $f(\mathcal{Q},\mathbb{T})$
gravity through a non-linear model, deriving cosmological implications consistent
with supernova data and the $\Lambda$CDM model. Agrawal et al. \cite{ii} examined an
extension of STG concerning late-time cosmic acceleration. Their work derives
dynamical parameters and validate non-singular matter bounce models through energy
conditions and stability analysis. Shiravand et al. \cite{jj} investigated
cosmological inflation within $f(\mathcal{Q},\mathbb{T})$ gravity, deriving modified
slow-roll parameters and spectral indices. Their findings demonstrate alignment with
Planck 2018 observational data by appropriately constraining free parameters. Tayde
et al. \cite{kk} analyzed the potential for wormholes within the framework of
$f(\mathcal{Q},\mathbb{T})$ gravity. Recent studies underline the versatility of
$f(\mathcal{Q}, \mathbb{T})$ in addressing cosmological and astrophysical problems
\cite{14a}-\cite{14a5}.

When a star runs out of nuclear fuel, it loses pressure, collapses
and forms new dense stars called compact stars. Baade and Zwicky
\cite{7-a} proposed that compact stars are formed as a result of
supernovae, a theory supported by the discovery of pulsars
\cite{7-b}. Pulsars are neutron stars that rotate and possess strong
magnetic field emitting the beams of electromagnetic radiation.
Examples like the pulsar star SAX J1748.9-2021 (PS) are crucial for
understanding neutron star physics, accretion processes, and matter
behavior in extreme gravitational and magnetic fields. They also
offer powerful insights into binary system evolution and the final
stages of stellar life cycles. In 1998, a sudden outburst led to the
first observation of the neutron star, PS in a dense star cluster.
Following this discovery, scientists identified its visible and
inactive forms \cite{7-bb}. In 2001 \cite{7-bbb}, 2005 \cite{7-bbbb}
and 2010 \cite{7-bbbbb}, more outbursts occurred, during which
several X-ray bursts were detected. Sixteen of these bursts observed
with the Rossi X-ray timing explorer that showed the signs of
powerful explosions on the neutron stars surface.

The fascinating properties and structures of neutron stars have
attracted considerable attention. Mak and Harko \cite{7-c} studied
the stability of pulsars by looking at energy limits and their
stable state through the speed of sound. Kramer et al. \cite{7-d}
highlighted the significance and potential of the double pulsar
system PSR J0737-3039A/B for performing precise tests of GR and
modified theories. Sharma et al. \cite{7-dd} analyzed the 2017
outburst of PS using the AstroSat telescope and discovered key
spectral and timing characteristics of the pulsar, including an
average spin frequency of 442.361098 Hz. Recently, the impacts of
$f(\mathcal{R})$ and $f(\mathcal{R},\mathbb{T})$ theories on PS are
discussed \cite{7-e, 7-e1}.

This paper aims to investigate the viable characteristics of
anisotropic PS within the framework of $f(\mathcal{Q}, \mathbb{T})$
theory. The structure of this paper is as follows. Section
\textbf{2} provides a comprehensive overview of $f(\mathcal{Q},
\mathbb{T})$ theory and its field equations. In section \textbf{3},
the field equations for a specific model $f(\mathcal{Q},
\mathbb{T})$ and the Krori-Barua (KB) ansatz are employed. We use
matching conditions to determine the unknown constants in the KB
ansatz. In sections \textbf{4} and \textbf{5}, we utilize
observational data of PS to find density, radial and tangential
pressures, anisotropy, mass-radius relation, redshift, the Zeldovich
condition, energy conditions, causality condition, adiabatic index
and Tolman-Oppenheimer-Volkoff (TOV) equation, EoS parameter and
compactness. We also assess stability of the model based on these
physical constraints. The final section presents our main results.

\section{Formalism of $f(\mathcal{Q}, \mathbb{T})$ Theory}

This section presents the basic framework of the modified
$f(\mathcal{Q}, \mathbb{T})$ theory, employing the variational
principle to derive the field equations. Weyl \cite{3-a} proposed an
extension of Riemannian geometry as a mathematical basis for
explaining gravitation in the context of GR. This modification
introduces a new vector field, $\Psi^{\alpha}$, which delineates
geometric properties of the Weyl geometry.  Weyl theory suggests
that a vector field shares similar mathematical properties with
electromagnetic potentials in physics, highlighting a strong
connection between gravitational and electromagnetic forces
\cite{21-a}. In Weyl geometry, when a vector of length $y$ is
transported along an infinitesimal path $\delta x^{\xi}$, its new
length is given by $\delta y = y \Psi_{\xi }\delta x^{\xi}$.

The connection in Weyl-Cartan space can be represented as
\begin{equation}\label{8}
{\hat\Gamma^{\lambda}_{\xi\varrho}}=\Gamma^{\lambda}_{\xi\varrho}
+\mathrm{C}^{\lambda}_{\xi\varrho}+\mathrm{L}^{\lambda}_{\xi\varrho},
\end{equation}
where $\Gamma^{\lambda}_{\xi\varrho}$ denotes the usual Christoffel
symbol, \(\mathrm{C}^{\lambda}_{\xi\varrho}\) represents the
contortion tensor, \(\mathrm{L}^{\lambda}_{\xi\varrho}\) indicates
the disformation tensor. The contortion tensor can be derived
\begin{equation}\label{11}
\mathrm{C}^{\lambda}_{\xi\varrho}=\hat\Gamma^{\lambda}_{[\xi\varrho]}
+g^{\lambda\phi}g_{\xi\kappa}\hat\Gamma^{\kappa}_{[\varrho\phi]}
+g^{\lambda\phi}g_{\varrho\kappa}\hat\Gamma^{\kappa}_{[\xi\phi]}.
\end{equation}
The disformation tensor can be obtained as
\begin{equation}\label{12}
\mathrm{L}^{\lambda}_{\xi\varrho}=\frac{1}{2}g^{\lambda\phi}(\mathcal{Q}_{\varrho\xi\phi}
+\mathcal{Q}_{\xi\varrho\phi}-\mathcal{Q}_{\lambda\xi\varrho}),
\end{equation}
where
\begin{equation}\label{13}
\mathcal{Q}_{\lambda\xi\varrho} = \nabla_{\lambda} g_{\xi\varrho}.
\end{equation}
The Weyl-Cartan torsion is described as
\begin{equation}\label{16}
T^{\lambda}_{\xi\varrho}=\frac{1}{2}(\hat{\Gamma}^{\lambda}_{\xi\varrho}
-\hat{\Gamma}^{\lambda}_{\xi\varrho}).
\end{equation}
The connection is defined by the disformation tensor as
\begin{align}\label{19}
\Gamma^{\lambda}_{\xi\varrho}&=-\mathrm{L}^{\lambda}_{\xi\varrho}.
\end{align}

The gravitational action in a non-covariant form is expressed as
\cite{4}
\begin{equation}\label{20}
S=\frac{1}{2\kappa} \int g^{\xi\varrho}(\Gamma^{\rho}_{\phi\xi}
\Gamma^{\phi}_{\varrho\rho} -\Gamma^{\rho}_{\phi\rho}
\Gamma^{\phi}_{\xi\varrho})\sqrt{-g}d^{4}x,
\end{equation}
where $\kappa$ signifies the coupling constant, which is chosen to
be one, $g$ indicates determinant of the metric tensor. Utilizing
the relation \eqref{19}, the action integral becomes
\begin{equation}\label{21}
S=-\frac{1}{2} \int g^{\xi\varrho}(\mathrm{L}^{\rho}_{\phi\xi}
\mathrm{L}^{\phi}_{\varrho\rho} - \mathrm{L}^{\rho}_{\phi\rho}
\mathrm{L}^{\phi}_{\xi\varrho}) \sqrt{-g} d^{4}x.
\end{equation}
This is known as the action of STG and is equivalent to the
Einstein-Hilbert action. There are several important differences
between the two gravitational paradigms. One notable feature is that
the disappearance of the curvature tensor in STG results in the
system appearing as a uniformly flat structure. Moreover,
gravitational effects are attributed to variations in the length of
a vector, rather than to the rotation of the angle between two
vectors during parallel transport.

Next, we examine an extension of the STG Lagrangian given by
Eq.(\ref{21}), which is formulated as follows
\begin{equation}\label{7}
S=\int\left[\frac{1}{2}f(\mathcal{Q})+L_{m}\right] \sqrt{-g}d^{4}x,
\end{equation}
where $L_{m}$ represents the matter Lagrangian. Now, we extend this
gravitational Lagrangian by adding the trace of the EMT to the
functional action as follows
\begin{equation}\label{22}
S=\int\bigg[\frac{1}{2}f(\mathcal{Q},
\mathbb{T})+L_{m}\bigg]\sqrt{-g}d^{4}x.
\end{equation}
Moreover
\begin{equation}\label{23}
\mathcal{Q}=-g^{\xi\varrho}(\mathrm{L}^{\rho}_{\mu\xi}\mathrm{L}^{\mu}_{\varrho\rho}
-\mathrm{L}^{\rho}_{\mu\rho}\mathrm{L}^{\mu}_{\xi\varrho}),
\end{equation}
where
\begin{equation}\label{24}
\mathrm{L}^{\rho}_{\mu\varpi}=-\frac{1}{2}g^{\rho\lambda}
(\nabla_{\varpi}g_{\mu\lambda}+\nabla_{\mu}g_{\lambda\varpi}
-\nabla_{\lambda}g_{\mu\varpi}).
\end{equation}
The traces of the non-metricity tensor are defined by the following
expressions
\begin{align}\label{25}
\mathcal{Q}_{\rho}= \mathcal{Q}^{~\xi}_{\rho~\xi}, \quad
\tilde{\mathcal{Q}}_{\rho}= \mathcal{Q}^{\xi}_{\rho\xi}.
\end{align}
The superpotential with respect to $\mathcal{Q}$ is expressed as
\begin{align}\label{26}
P^{\rho}_{\xi\varrho}&=-\frac{1}{2}\mathrm{L}^{\rho}_{\xi\varrho}
+\frac{1}{4}(\mathcal{Q}^{\rho}-\tilde{\mathcal{Q}}^{\rho})
g_{\xi\varrho}- \frac{1}{4} \delta ^{\rho}_{~[\xi
\mathcal{Q}_{\varrho}]}.
\end{align}
Furthermore, the expression for $\mathcal{Q}$ obtained through the superpotential can
be written as follows \cite{5-g}
\begin{align}\label{27}
\mathcal{Q}=-\mathcal{Q}_{\rho\xi\varrho}P^{\rho\xi\varrho}
=-\frac{1}{4}(-\mathcal{Q}^{\rho\xi\varrho}\mathcal{Q}_{\rho\varrho\xi}
+2\mathcal{Q}^{\rho\varrho\xi}\mathcal{Q}_{\xi\rho\varrho}
-2\mathcal{Q}^{\varrho}\tilde{\mathcal{Q}}_{\varrho}
+\mathcal{Q}^{\varrho}\mathcal{Q}_{\varrho}).
\end{align}
The field equations are obtained by taking the variation of $S$ with
respect to the metric tensor and equating it to zero as follows
\begin{align}\nonumber
\delta S&=0=\int \frac{1}{2}\delta
[f(\mathcal{Q},\mathbb{T})\sqrt{-g}]+\delta[L_{m}\sqrt{-g}]d^{4}x
\\\label{28} 0&=\int \frac{1}{2}\bigg( \frac{-1}{2} f g_{\xi\varrho}
\sqrt{-g} \delta g^{\xi\varrho} + f_{\mathcal{Q}} \sqrt{-g} \delta
\mathcal{Q} + f_{\mathbb{T}} \sqrt{-g} \delta
\mathbb{T}\bigg)-\frac{1}{2} \mathbb{T}_{\xi\varrho} \sqrt{-g}
\delta g^{\xi\varrho}d^ {4}x.
\end{align}
Additionally, we define
\begin{align}\label{29}
\mathbb{T}_{\xi\varrho} &= \frac{-2}{\sqrt{-g}} \frac{\delta
(\sqrt{-g} L_{m})}{\delta g^{\xi\varrho}}, \quad
\Theta_{\xi\varrho}= g^{\rho\mu} \frac{\delta
\mathbb{T}_{\rho\mu}}{\delta g^{\xi\varrho}},
\end{align}
which implies that $ \delta \mathbb{T}=\delta
(\mathbb{T}_{\xi\varrho}g^{\xi\varrho})=(\mathbb{T}_{\xi\varrho}+
\Theta_{\xi\varrho})\delta g^{\xi\varrho}$. Inserting the
aforementioned factors in Eq.\eqref{28}, we have
\begin{eqnarray}\nonumber
\delta S =0&=&\int \frac{1}{2}\bigg\{\frac{-1}{2}f
g_{\xi\varrho}\sqrt{-g} \delta g^{\xi\varrho} +
f_{\mathbb{T}}(\mathbb{T}_{\xi\varrho}+
\Theta_{\xi\varrho})\sqrt{-g} \delta g^{\xi\varrho}
\\\nonumber
&-&f_{\mathcal{Q}} \sqrt{-g} (P_{\xi\rho\mu}
\mathcal{Q}_{\varrho}~^{\rho\mu}- 2\mathcal{Q}^{\rho\varrho}~_{\xi}
P_{\rho\mu\varrho}) \delta g^{\xi\varrho}+2f_{\mathcal{Q}} \sqrt{-g}
P_{\rho\xi\varrho} \nabla^{\rho} \delta g^{\xi\varrho}
\\\label{30}
&+&2f_{\mathcal{Q}}\sqrt{-g}P_{\rho\xi\varrho} \nabla^{\rho} \delta
g^{\xi\varrho} \bigg\}- \frac{1}{2} \mathbb{T}_{\xi\varrho}\sqrt{-g}
\delta g^{\xi\varrho}d^ {4}x.
\end{eqnarray}
When integrating the term $ 2f_{\mathcal{Q}} \sqrt{-g}
P_{\rho\xi\varrho}\nabla^{\rho}\delta g^{\xi\varrho}$ and applying
the boundary conditions, the resulting expression is $ -2
\nabla^{\rho} (f_{\mathcal{Q}} \sqrt{-g} P_{\rho\xi\varrho})\delta
g^{\xi\varrho}$. The partial derivatives with respect to
$\mathcal{Q}$ and $\mathbb{T}$ are represented by $f_{\mathcal{Q}}$
and $f_{\mathbb{T}}$, respectively. Finally, we can write the field
equations as follows
\begin{eqnarray}\nonumber
\mathbb{T}_{\xi\varrho}&=&\frac{-2}{\sqrt{-g}} \nabla_{\rho}
(f_{Q}\sqrt{-g} P^{\rho}_{\xi\varrho})- \frac{1}{2} f g_{\xi\varrho}
+ f_{\mathbb{T}} (\mathbb{T}_{\xi\varrho} +
\Theta_{\xi\varrho})\\\label{31}
&-&f_{\mathcal{Q}} (P_{\xi\rho\mu}
\mathcal{Q}_{\varrho}~^{\rho\mu} -2\mathcal{Q}^{\rho\mu}~_{\xi}
P_{\rho\mu\varrho}).
\end{eqnarray}

To study the internal structure of a star in equilibrium, we examine
a spherically symmetric system. The line element is given as
\begin{equation}\label{a1}
ds^{2}=-e^{\alpha(r)}dt^{2}+e^{\beta(r)}dr^{2}+r^{2}(d\theta^{2}+\sin^{2}d\phi^{2}).
\end{equation}
We consider the stellar matter distribution to be an anisotropic,
which can be characterized by
\begin{equation}\label{a2}
{\mathbb{T}}_{\xi\varrho}=(\rho+P_{t})u_{\xi}u_{\varrho}+P_{t}g_{\xi\varrho}-\sigma
k_{\xi}k_{\varrho},
\end{equation}
here, $u_{\xi}$ and $u_{\varrho}$ represent the four-velocity of the
fluid, adhering to the normalization condition $u^\xi u_\xi = -1$.
Additionally, $k_{\xi}$ is a unit radial four-vector that satisfies
the condition $ k^{\xi} k_{\xi} = 1$. In this context, the energy
density is $\rho$, the tangential pressure is  $P_{t}$, the radial
pressure is $P_{r}$ and the anisotropy factor is $\sigma$ defined as
$(P_{t} - P_{r})$. Based on the given metric, we can write $u^{\xi}$
as $e^{-\alpha} \delta^\xi_{t}$, and $k^\xi$ as $e^{-\beta}
\delta^\xi_{r}$. Thus, the trace of the EMT in Eq.\eqref{a1} can be
represented as
\begin{equation}\label{a3}
\tilde{\mathbb{T}}=-\rho+3P_{r}+2\sigma.
\end{equation}
Taking the divergence of Eq.\eqref{a2} results in the conservation
law for energy and momentum, expressed as
\begin{equation}\label{a4}
\nabla_{\varrho}\tilde{\mathbb{T}}^{~\varrho}_{1}=(\rho+P_{r})
\alpha^{\prime}+P_{r}^{\prime}-\frac{1}{r}\sigma,
\end{equation}
where prime is the derivative with respect to $r$. The non-zero
parts of the field equations \eqref{31} are
\begin{align}\nonumber
\rho&=\frac{1}{2 r^2 e^{\beta (r)}}\bigg[f_{\mathcal{Q}} \big(\big(e^{\beta
(r)}-1\big) \big(r \alpha '(r)+2\big)+r \big(e^{\beta (r)}+1\big) \beta '(r)\big)
\\\label{1-a}
&+2f_{\mathcal{Q}\mathcal{Q}} r\big(e^{\beta (r)}-1\big) Q'(r)+f r^2 e^{\beta
(r)}\bigg]-\frac{1}{3}f_{\mathbb{T}} (P_{r}+2P_{t}+3 \rho ),
\\\nonumber
P_{r}&=\frac{1}{2 r^2 e^{\beta (r)}}\bigg[f_{\mathcal{Q}} \big(\big(e^{\beta
(r)}-1\big) \big(r \alpha '(r)+r \beta '(r)+2\big)-2 r \alpha '(r)\big)
\\\label{1-b}
&+2 f_{\mathcal{Q}\mathcal{Q}} r \big(e^{\beta (r)}-1\big)
Q'(r)\bigg]+\frac{2}{3}f_{\mathbb{T}} (P_{t}-P_{r})+f r^2 e^{\beta (r)},
\\\nonumber
P_{t}&=\frac{1}{3}f_{\mathbb{T}}(P_{r}-P_{t})-\frac{1}{4 r e^{\beta
(r)}}\bigg[f_{\mathcal{Q}} \big(-2 r \alpha ''(r)+\beta '(r) \big(r \alpha '(r)+2
e^{\beta (r)}\big)
\\\label{1-c}
&+2 \big(e^{\beta (r)}-2\big)\alpha '(r)-r \alpha '(r)^2\big)-2
f_{\mathcal{Q}\mathcal{Q}} r Q'(r) \alpha '(r)+2 f r e^{\beta (r)}\bigg].
\end{align}

\section{The $f(\mathcal{Q}, \mathbb{T})$ Gravity Model}

We now examine the influence of $f(\mathcal{Q}, \mathbb{T})$ on the
geometry of PS. For this purpose, we utilize the specific model of
$f(\mathcal{Q}, \mathbb{T})$ as \cite{10-b}.
\begin{align}\label{A}
f(\mathcal{Q}, \mathbb{T})&=\zeta\mathcal{Q} + \eta\mathbb{T},
\end{align}
where $\zeta$ and $\eta$ are arbitrary constants with
$f_\mathcal{Q}=\zeta,~ f_{\mathcal{QQ}}=0,~ f_\mathcal{T}=\eta,~
f_{\mathcal{TT}}=0$. This cosmological model is extensively employed
in the literature \cite{10-c}. Substituting these values into
Eqs.\eqref{1-a}-\eqref{1-c}, we obtain
\begin{align}\nonumber
\rho&=\frac{1}{2 r^2 \big(\eta  \big(-8 \eta +2 (4 \eta +3) r^2
e^{\beta (r)}-15\big)-6\big)}\big[\zeta  e^{-\beta (r)} \big(r
\big(\eta \alpha '(r) \big(\big(e^{\beta (r)}-1\big)
\\\nonumber
&\times\big(2 r^2 e^{\beta (r)}-1\big)+r \beta '(r)\big)+\big(\eta
\big(e^{\beta (r)} \big(2 r^2 \big(e^{\beta
(r)}+5\big)-1\big)-9\big)-12\big) \beta '(r)
\\\label{a-1}
& -2 \eta r \alpha ''(r)+\eta  (-r) \alpha '(r)^2\big)+2
\big(e^{\beta (r)}-1\big) \big(\eta  \big(6 r^2 e^{\beta
(r)}-7\big)-6\big)\big)\big],
\\\nonumber
P_{r}&=\frac{1}{2 (\eta +1) r^2 \big(\eta  \big(-8 \eta +2 (4 \eta
+3) r^2 e^{\beta (r)}-15\big)-6\big)}\bigg[\zeta  e^{-\beta (r)}
\big(r \big(-2 \eta  r \alpha ''(r)
\\\nonumber
&\times \big(3 \eta +2 (4 \eta +3) r^2 e^{\beta (r)}+2\big)+\alpha
'(r) \big(\eta  (27 \eta +47)+\eta r \big(3 \eta +2 (4 \eta +3) r^2
\\\nonumber
&\times e^{\beta (r)}+2\big) \beta '(r)+2 (\eta +1) (11 \eta +6) r^2
e^{2 \beta (r)}-e^{\beta (r)} \big((\eta +1) (11 \eta +6)
\\\nonumber
&+2 (\eta (19 \eta+23) +6) r^2\big)+18\big)+\eta (-r) \alpha '(r)^2
\big(3 \eta +2 (4 \eta +3) r^2 e^{\beta (r)}+2\big)
\\\nonumber
&+\big(\eta (13 \eta +21)+2 (\eta +1) (11 \eta +6) r^2 e^{2 \beta
(r)} -e^{\beta (r)} \big((\eta +1) (11 \eta +6)+2
\\\nonumber
&\times (\eta +2)(\eta +3) r^2\big)+6\big) \beta '(r)\big)+2
\big(e^{\beta (r)}-1\big) \big(\eta  \big(-13 \eta +2 (\eta +3) r^2
\\\label{a-2}
&\times e^{\beta (r)}-17\big)-6\big)\big)\bigg],
\\\nonumber
P_{t}&=\frac{1}{4 (\eta +1) r^2 \big(\eta  \big(-8 \eta +2 (4 \eta
+3) r^2 e^{\beta (r)}-15\big)-6\big)}\bigg[\zeta e^{-\beta (r)}
\big(r \big(2 \big(r \alpha ''(r)
\\\nonumber
&\times\big(\eta  \big(-6 \eta +2 (4 \eta +3) r^2 e^{\beta
(r)}-13\big)-6\big)+\big(\eta \big(\eta +e^{\beta (r)} \big(\eta -2
r^2 \big(\eta +(\eta +1)
\\\nonumber
&\times e^{\beta (r)}+2\big)+1\big)+6\big)+6\big) \beta '(r)\big)
+\alpha '(r) \big(r \big(\eta  \big(6 \eta -2 (4 \eta +3) r^2
e^{\beta (r)}+13\big)
\\\nonumber
&+6\big)\beta '(r)+2 \eta \big(-9 \eta +e^{\beta (r)} \big(\eta -2
r^2 \big(-5 \eta +(\eta +1) e^{\beta (r)}-4\big)+1\big)-16\big)
\\\nonumber
&-12\big)+r \alpha '(r)^2 \big(\eta \big(-6 \eta +2 (4 \eta +3) r^2
e^{\beta (r)}-13\big)-6\big)\big)+4 \eta \big(e^{\beta (r)}-1\big)
\\\label{a-3}
&\times\big(\eta\big(2 r^2 e^{\beta (r)}-1\big)-2\big)\big)\bigg].
\end{align}
We make some reasonable assumptions about the metric potentials
$\alpha(r)$ and $\beta(r)$ to study stellar evolution. We use the KB
spacetime to investigate stellar models within the framework of
modified gravity theory.

\subsection{The Krori-Barua Metric Potentials}

This is a well-known ansatz used in GR to model the interior of
stars. It provides a specific form of the metric potentials that
satisfy the Einstein field equations under certain conditions. The
KB metric is particularly useful for studying the structure and
properties of highly dense astrophysical objects such as PS. The
metric potentials of KB are defined as \cite{10-d}
\begin{equation}\label{4-a}
\alpha(r)=A_{1}\bigg(\frac{r}{R}\bigg)^{2} + a_{1},~~~\beta(r) =
a\bigg(\frac{r}{R}\bigg)^{2},
\end{equation}
where $R$ represents the radius of the star and $A_{1}, a_{1}$ and
$a$ are arbitrary constants and can be found by using junction
conditions. It helps in understanding how different theoretical
frameworks such as $f(\mathcal{Q},\mathbb{T})$ gravity affect the
properties of these stars.

The KB metric is particularly advantageous within the
$f(\mathcal{Q}, \mathbb{T})$ gravity framework because of its smooth
and non-singular behavior at both the stellar core $r = 0$ and
surface $r = R$. This ensures that physical parameters, such as
density and pressure, remain well-behaved throughout the stellar
interior, avoiding singularities that could compromise the model
physical validity. Additionally, the KB metric is robust and
flexible, making it adaptable to modified gravity frameworks like
$f(\mathcal{Q}, \mathbb{T})$, where geometric modifications arise
from non-metricity and the trace of the EMT. Its use is further
justified by its ability to incorporate observational mass-radius
constraints, such as those from PS, thereby bridging theoretical
modeling with real-world astrophysical data.

\subsection{Matching Conditions}

Matching conditions are essential in the study of stellar structures
and GR. These conditions ensure that the interior solution of a
star, described by its matter distribution and gravitational field,
smoothly connects with the exterior vacuum solution, typically
represented by the Schwarzschild solution, given by
\begin{equation}\label{c-1}
ds^{2}=-\bigg(1-\frac{2GM}{c^{2}r}\bigg)c^{2}dt^{2}
+\bigg(1-\frac{2GM}{c^{2}r}\bigg)^{-1}dr^{2}+r^{2}(d\theta+\sin^{2}\theta
d\phi^{2}),
\end{equation}
where $M$ denotes the mass of the star, $G$ is the gravitational
constant and $c$ is the speed of light. The continuity of the metric
coefficients for the metrics \eqref{a1} and \eqref{c-1} at the
surface boundary $(r = R)$ results in
\begin{align}\label{cc-1}
g_{tt}&=e^{A_{1}\big(\frac{r}{R}\big)^{2} +
a_{1}}=-\bigg(1-\frac{2GM}{c^{2}r}\bigg)c^{2},
\\\label{cc-2}
g_{rr}&=e^{a\big(\frac{r}{R}\big)^{2}}=\bigg(1-\frac{2GM}{c^{2}r}\bigg)^{-1}.
\end{align}
Applying the matching conditions, we have
\begin{equation}\label{c-3}
\alpha(r=R)=\ln(-u+1),\quad\beta(r=R)=-\ln(-u+1),\quad P_{r}(r=R)=0,
\end{equation}
where $u$ denotes compactness, defined as
\begin{equation}\label{c-4}
u=\frac{2GM}{c^{2}R},
\end{equation}
here, we use the gravitational units such that $c=G=1$.

Having established the rationale for the KB metric, we now discuss
the significance of the boundary condition $P_r = 0$ at $r = R$.
Physically, at the stellar surface, there is no external pressure
acting on the star, as the surrounding region is a vacuum described
by the Schwarzschild solution. This necessitates setting the $P_r =
0$ at $r = R$, ensuring a smooth and natural transition between the
star interior solution (with anisotropic matter) and the exterior
vacuum solution. Mathematically, this condition guarantees the
continuity of the metric coefficients $g_{tt}$ and $g_{rr}$ and
their derivatives at the stellar boundary. Such continuity is
essential for satisfying the junction conditions required in GR and
its extensions, including $f(\mathcal{Q}, \mathbb{T})$ gravity. The
absence of any pressure discontinuity eliminates non-physical
behavior at the surface, which could otherwise destabilize the star.

The boundary condition, $P_{r}=0$, also has a direct impact on the
model stability and accuracy. It ensures hydrostatic equilibrium,
where the outward pressure gradient balances the inward
gravitational pull, preventing matter from escaping and maintaining
structural stability. In terms of accuracy, this condition is
critical for determining free parameters, such as the metric
coefficients, which define the star internal structure. It enables
the derivation of a physically realistic mass-radius relation and
compactness, aligning the theoretical predictions with observational
data like PS. Without these boundary conditions, the model would
exhibit pressure inconsistencies, unstable behavior, and unrealistic
predictions, ultimately undermining its physical validity.
Table~\ref{table:units} ensures numerical consistency and address
concerns regarding units that explicitly states all the physical
quantities used in our analysis along with their corresponding
units.
\begin{table}
\centering \caption{Physical quantities and their units used in the
analysis.}\vspace{0.5cm}
\begin{tabular}{|c|c|c|}
\hline \textbf{Variable} & \textbf{Physical Meaning}        &
\textbf{Units}
\\ \hline $\rho$            & Density                         & $g/cm^{3}$
\\ \hline $P_{r}, P_{t}$    & Radial/Tangential pressure      & $dyn/cm^{2}$
\\ \hline $u$               & Compactness ($u = \frac{M}{R}$) & Dimensionless
\\ \hline $v_r^2, v_t^2$    & Squared sound speeds (Radial/Tangential) &Dimensionless
\\ \hline $Z_s$             & Gravitational redshift& Dimensionless
\\ \hline $\Gamma$          & Adiabatic index& Dimensionless
\\ \hline
\end{tabular}
\label{table:units}
\end{table}

\section{Stability Analysis from SAX J 1748.9-2021 Observations}

In this section, we use observational data focusing on the mass and
radius of PS within the framework of $f(\mathcal{Q}, \mathbb{T})$
gravity. The observational data for PS was obtained from
spectroscopic measurements during thermonuclear bursts, as reported
in prior studies using the Rossi X-ray timing explorer. This dataset
provides high-precision measurements of the pulsar's mass $M = 1.81
\pm 0.3 M_\odot$ and radius $(R = 11.7 \pm 1.7)$km, making this
pulsar an ideal candidate for testing modified gravity models. The
unique characteristics of this PS, such as its rapid spin frequency
$(442.361098)$Hz and compact configuration, further justify its
selection for this study. These properties offer an excellent
opportunity to examine strong-field deviations from GR and validate
the predictions of the $f(\mathcal{Q}, \mathbb{T})$ framework.
Additionally, we outline the criteria for selecting PS as the focus
of our study, emphasizing the availability of accurate and reliable
observational data that align well with the theoretical framework of
$f(\mathcal{Q}, \mathbb{T})$ gravity. To further strengthen the
reliability of our results, we have conducted a systematic
uncertainty analysis by propagating uncertainties in the
observational data through the field equations.

\subsection{Material Component}

In the study of stellar models, key physical quantities such as
$\rho$, $P_{r}$, $P_{t}$ and anisotropy play crucial roles in
determining the behavior and stability of the star. They provide
insights into the internal structure and dynamics of the material
component of the star. The quantities $\rho$, $P_{r}$ and $P_{t}$
are derived from the model's field equations using a combination of
analytical and numerical methods. Initially, the field equations are
solved analytically and then mathematical software, such as
Mathematica, is employed to obtain simplified expressions for these
quantities. This approach helps us to manage the algebraic
complexity effectively and derives the simplified forms of $\rho$,
$P_{r}$ and $P_{t}$. Once these quantities are determined through
the field equations, we have applied a linear model and utilized the
KB metric to further simplify the equations. By following these
steps, we have successfully derived the following quantities
\begin{align}\nonumber
\rho&=\bigg[\zeta  e^{-\frac{a r^2}{R^2}} \big(-2 A_{1}^2 \eta r^4+2 A_{1} a \eta
r^4+2 \eta r^2 R^2 e^{\frac{2 a r^2}{R^2}} \big(r^2 (A_{1}+a) +3 R^2\big)
\\\nonumber
&-R^2 e^{\frac{a r^2}{R^2}}\big(\eta  r^2 \big(2 r^2 (A_{1}-5 a)+A_{1}+a\big)+R^2
\big(\eta \big(6 r^2+7\big)+6\big)\big)-A_{1}
\\\nonumber
&\times\eta r^2 R^2-9 a \eta  r^2 R^2-12 a r^2 R^2+7 \eta R^4+6
R^4\big)\bigg]\bigg[r^2 R^4 \big(\eta \big(2 (4 \eta +3) r^2 e^{\frac{a r^2}{R^2}}
\\\label{5a}
&-8\eta -15\big)-6\big)\bigg]^{-1},
\\\nonumber
P_{r}&=\bigg[\zeta  e^{-\frac{a r^2}{R^2}} \big(\frac{1}{R^4}\bigg[2 r^2
\big(-e^{\frac{a r^2}{R^2}} \big(4 A_{1}^2 \eta  (4 \eta +3) r^4+A_{1} \big(R^2
\big((\eta +1) (11 \eta +6)
\\\nonumber
&+2 (\eta (27 \eta +29)+6) r^2\big)-4 a \eta  (4 \eta +3) r^4\big)+a R^2 \big((\eta
+1) (11 \eta +6)+2
\\\nonumber
&\times(\eta +2)(\eta +3)r^2\big)\big)-6 A_{1}^2 \eta ^2 r^2-4 A_{1}^2 \eta r^2+6
A_{1} a \eta ^2 r^2+4 A_{1} a \eta  r^2+2
\\\nonumber
&\times(\eta +1)(11 \eta +6) r^2R^2 (A_{1}+a) e^{\frac{2 a r^2}{R^2}}+21 A_{1} \eta
^2 R^2+43 A_{1} \eta R^2+18 A_{1} R^2
\\\nonumber
&+13 a \eta ^2 R^2+a \eta  R^2+6 a R^2\big)\bigg]+2 \big(e^{\frac{a r^2}{R^2}}-1\big)
\big(\eta \big(2 (\eta +3) r^2 e^{\frac{a r^2}{R^2}}-13 \eta \big)\big)\big)\bigg]
\\\label{5b}
&\times\bigg[2 (\eta +1) r^2 \big(\eta \big(2 (4 \eta +3) r^2 e^{\frac{a r^2}{R^2}}-8
\eta -15\big)-6\big)\bigg]^{-1},
\\\nonumber P_{t}&=\bigg[\zeta  e^{-\frac{a r^2}{R^2}}
\big(A_{1}^2 r^4 \big(\eta  \big(2 (4 \eta +3) r^2 e^{\frac{a r^2}{R^2}}-6 \eta
-13\big)-6\big)+A_{1} r^2 \big(a r^2 \big(\eta \big(-2
\\\nonumber&
+(4 \eta +3) r^2 e^{\frac{a r^2}{R^2}}6 \eta +13\big)+6\big)+R^2 \big(\eta
\big(e^{\frac{a r^2}{R^2}} \big(2 r^2 \big(-(\eta +1)e^{\frac{a r^2}{R^2}}+9 \eta
+7\big)
\\\nonumber
&+\eta +1\big)-15 \eta -29\big)-12\big)\big)+R^2 \big(a r^2 \big(\eta \big(e^{\frac{a
r^2}{R^2}}\big(2 r^2 \big(-(\eta +1) e^{\frac{a r^2}{R^2}}-\eta -2\big)
\\\nonumber
&+\eta +1\big)+\eta +6\big)+6\big) +\eta  R^2 \big(e^{\frac{a r^2}{R^2}}-1\big)
\big(2 \eta r^2 e^{\frac{a r^2}{R^2}}-\eta -2\big)\big)\big)\bigg]
\\\label{5c}
&\times \bigg[r^2 R^4 \big(\eta \big(2 (\eta +1) (4 \eta +3) r^2 e^{\frac{a
r^2}{R^2}}-\eta (8 \eta +23)-21\big)-6\big)\bigg]^{-1}.
\end{align}
\begin{figure}
\epsfig{file=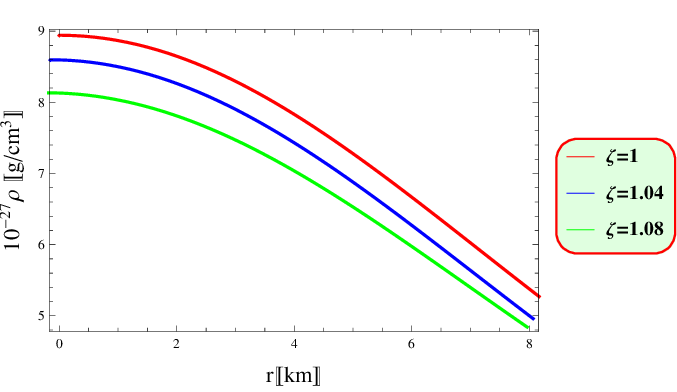,width=.52\linewidth}
\epsfig{file=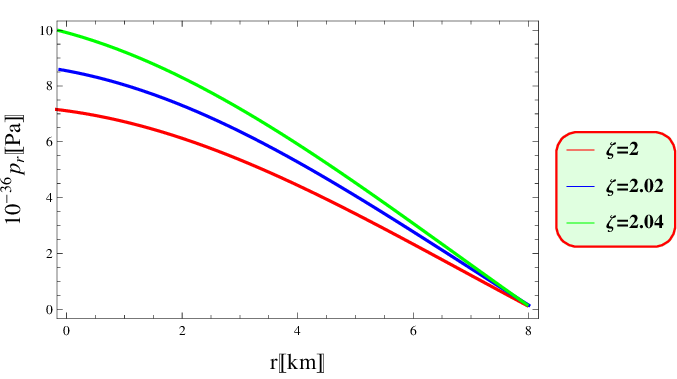,width=.52\linewidth} \epsfig{file=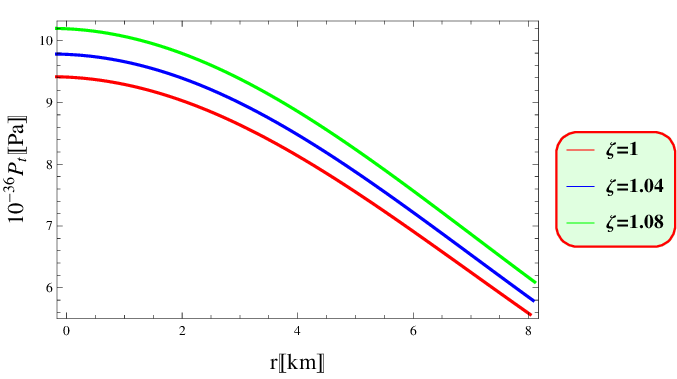,
width=.52\linewidth} \caption{Plots of $\rho$, $P_{r}$, $P_{t}$
versus $r$.}
\end{figure}

The constants are set as $A_{1}=2,~ a_{1}=0,~ a=-2$ and various
values of $\zeta=1, 1.04$ and $1.08$. Throughout the paper these
values are applied consistently to derive and evaluate our equations
and results. To further strengthen the reliability of our results,
we performed a systematic uncertainty analysis by propagating the
uncertainties in the radius through the field equations. We have
analyzed the impact of these uncertainties on key physical
parameters, such as $\rho, P_{r}, P_{t}$ and compactness. The
results confirmed that these parameters remain stable and within
physically expected bounds, validating the robustness of our model.
Additionally, we have conducted a sensitivity analysis by varying
the parameter $\zeta$ and studying its effect on the results. We
have observed a consistent increasing trend in the graphs for three
different values of $\zeta$. The parameter values were carefully
fixed to ensure smooth and physically consistent behavior in the
graphs. Testing lower values, such as $R = 10$km, led to deviations
from the expected behavior, reinforcing the choice of $R = 13.4$km
for producing meaningful and consistent results. We have employed
Mathematica for solving the field equations. Figure \textbf{1}
presents the plots of $\rho$, $P_{r}$ and $P_{t}$ as functions of
$r$, showing that these physical characteristics are peak at the
center and decrease outward indicating a highly compact profile for
the proposed PS. Additionally, the $P_{r}$ within PS consistently
decreases as $r$ increases.
\begin{figure}\center
\epsfig{file=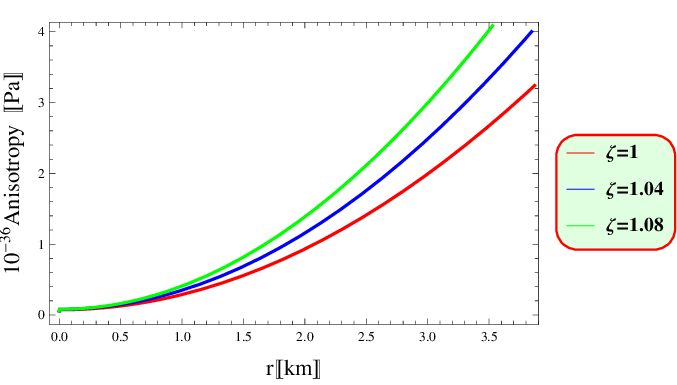,width=.55\linewidth}\caption{Plot of $\sigma$ against $r$.}
\end{figure}

Anisotropic $\sigma$ pressure refers to the condition where the
pressure within a material or system is not uniform in all
directions. This directional dependence of pressure can
significantly influence the behavior and stability of physical
systems, such as stars, where anisotropic pressure can affect their
structural integrity and evolution \cite{10-f}. The anisotropy
$(\sigma = P_{t} - P_{r})$ leads to
\begin{align}\nonumber
\sigma &=\frac{1}{r^2 R^4 \big(\eta  \big(2 (\eta +1) (4 \eta +3) r^2 e^{\frac{A_{1}
r^2}{R^2}}-\eta  (8 \eta +23)-21\big)-6\big)}\bigg[\zeta e^{-\frac{A_{1} r^2}{R^2}}
\big(a^2 r^4
\\\nonumber
&\times \big(\eta \big(2 (4 \eta +3) r^2 e^{\frac{A_{1} r^2}{R^2}}-6
\eta -13\big)-6\big)+a r^2 \big(A_{1} r^2 \big(\eta  \big(-2 (4 \eta
+3) r^2 e^{\frac{A_{1} r^2}{R^2}}
\\\nonumber
&+6 \eta +13\big)+6\big)+R^2 \big(\eta \big(e^{\frac{A_{1}
r^2}{R^2}} \big(2 r^2 \big(-(\eta +1) e^{\frac{A_{1} r^2}{R^2}}+9
\eta +7\big)+\eta +1\big)
\\\nonumber
&-15 \eta -29\big)-12\big)\big)+R^2 \big(A_{1} r^2 \big(\eta
\big(e^{\frac{A_{1} r^2}{R^2}} \big(2 r^2 \big(-(\eta +1)
e^{\frac{A_{1} r^2}{R^2}}-\eta -2\big)
\\\nonumber
&+\eta +1\big)+\eta +6\big)+6\big)+\eta  R^2 \big(e^{\frac{A_{1}
r^2}{R^2}}-1\big) \big(2 \eta  r^2 e^{\frac{A_{1} r^2}{R^2}}-\eta
-2\big)\big)\big)\bigg]
\\\nonumber
-&\bigg(2 (\eta +1) r^2 \big(\eta \big(2 (4 \eta +3) r^2
e^{\frac{A_{1} r^2}{R^2}}-8 \eta
-15\big)-6\big)\bigg)^{-1}\bigg[\zeta e^{-\frac{A_{1} r^2}{R^2}}
\bigg(\frac{1}{R^4}
\\\nonumber
&\times\bigg\{2 r^2 \big(-e^{\frac{A_{1} r^2}{R^2}} \big(4 a^2 \eta
(4 \eta +3) r^4+a \big(R^2 \big((\eta +1) (11 \eta +6)+2 (\eta  (27
\eta +29)
\\\nonumber
&+6) r^2\big) -4 A_{1} \eta  (4 \eta +3) r^4\big) +A_{1} R^2
\big((\eta +1) (11 \eta +6)+2 (\eta +2) (\eta +3) r^2\big)\big)
\\\nonumber
&-6 a^2 \eta ^2 r^2-4 a^2 \eta r^2+6 a A_{1} \eta ^2 r^2+4 a A_{1}
\eta  r^2+2 (\eta +1) (11 \eta +6) r^2 R^2 (a+A_{1})
\\\nonumber
&\times e^{\frac{2 A_{1} r^2}{R^2}}+21 a \eta ^2 R^2+43 a \eta
R^2+18 a R^2+13 A_{1} \eta ^2 R^2+21 A_{1} \eta R^2+6 A_{1}
R^2\big)\bigg\}
\\\nonumber
&+2 \big(e^{\frac{A_{1} r^2}{R^2}}-1\big)\times \big(\eta \big(2
(\eta +3) r^2 e^{\frac{A_{1} r^2}{R^2}}-13 \eta
-17\big)-6\big)\big)\bigg].
\end{align}
Figure \textbf{2} shows that the anisotropy begins at zero in the
core and steadily increases towards the surface of the star. It is
well-known that the positive anisotropy indicates stability
\cite{10-gg}, hence the considered PS is stable.

It is crucial to present numerical values for the physical
properties of the PS as predicted by the current model. With $\zeta
= 1.08$, $\rho_{core} \approx 8.93 \times 10^{27}$ g/cm$^{3}$,
$P_{r(core)} \approx 9.69 \times 10^{36}$ dyn/cm$^{2}$ and
$P_{t(core)} \approx 10.18 \times 10^{36}$ dyn/cm$^{2}$. At the
star's boundary, $\rho_{I}$ is $5.69 \times 10^{27}$ g/cm$^{3}$,
which is about 1.8 times the core density. Here $P_{r(r=R)} = 0$
dyn/cm$^{2}$ and $P_{t(r=R)} \approx 6.12 \times 10^{36}$
dyn/cm$^{2}$. The core values are $\rho_{core} \approx 8.57 \times
10^{27}$ g/cm$^{3}$, $P_{r(core)} \approx 8.18\times 10^{36}$
dyn/cm$^{2}$ and $P_{t(core)} \approx 9.76 \times 10^{36}$
dyn/cm$^{2}$ with $\zeta = 1.04$. At the star's boundary, $\rho_{I}$
is $5.18 \times 10^{27}$ g/cm$^{3}$, which is about 1.8 times the
core density. The radial/tangential pressures are $P_{r(r=R)} = 0$
dyn/cm$^{2}$ and $P_{t(r=R)} \approx 5.75 \times 10^{36}$
dyn/cm$^{2}$. The core values for $\zeta = 1$, $\rho_{core} \approx
8.031 \times 10^{27}$ g/cm$^{3}$, $P_{r(core)} \approx 7.025\times
10^{36}$ dyn/cm$^{2}$ and $P_{t(core)} \approx 9.40 \times 10^{36}$
dyn/cm$^{2}$. At the star's boundary, $\rho_{I}$ is $4.806 \times
10^{27}$ g/cm$^{3}$, which is about 1.8 times the core density,
while $P_{r(r=R)} = 0$ dyn/cm$^{2}$ and $P_{t(r=R)} \approx 5.51
\times 10^{36}$ dyn/cm$^{2}$. Table \ref{tab:comparison} provides
comparison of $\rho_{core}$, $P_{r(core)}$ and $P_{r(core)}$ under
different parameter values of $\zeta$.
\begin{table}
\centering \caption{Comparison of $\rho_{core}$, $P_{r(core)}$ and
$P_{r(core)}$ under different parameter values of $\zeta$.}
\label{tab:comparison} \vspace{0.5cm}
\begin{tabular}{|c|c|c|c|}
\hline $\zeta$ & $\rho_{core}~(10^{27}~g/cm^{3})$&$P_{r(core)}~(10^{36}~dyn/cm^{2})$
& $P_{t(core)}~(10^{36}~ dyn/cm^{2})$
\\ \hline 1.00 & 8.03 & 7.03 & 9.40
\\ \hline 1.04 & 8.57 & 8.18& 9.76
\\ \hline 1.08 & 8.93& 9.69& 10.18
\\ \hline
\end{tabular}
\end{table}

\subsection{Limit on the Mass-Radius Relation}

The mass of the PS is stated as \cite{10-g}
\begin{equation}\nonumber
M(r)= 4 \pi  \int_0^r \chi^2 \rho (\chi) d\chi.
\end{equation}
The numerical solution of this equation, assessing the reliability
of the current model, is shown in Figure \textbf{3}. The graph
clearly demonstrates that the mass increases steadily and uniformly
with the star's radius, hence satisfies the physical requirement
that mass is always positively increasing function.
\begin{figure}\center
\epsfig{file=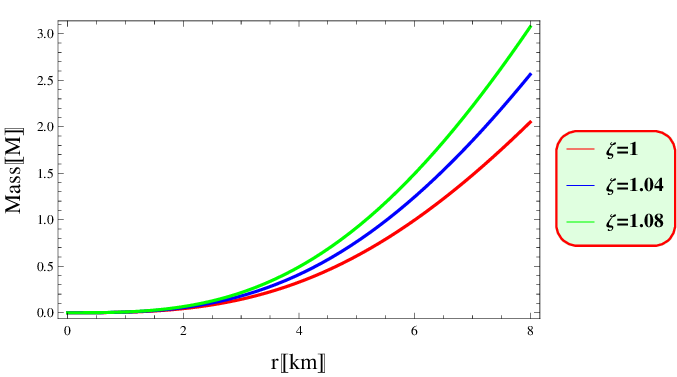,width=.55\linewidth}\caption{Plot of mass-radius against $r$.}
\end{figure}

\subsection{The Geometric Component}

The gravitational redshift, denoted by $Z_{s}(r)$, describes the
change in wavelength (or frequency) of light or other
electromagnetic radiation as it escapes from the gravitational
field. According to GR, this effect intensifies in stronger
gravitational field, particularly around massive objects like stars
or black holes. This is defined as follows
\begin{equation}\label{ccc}
Z_{s}(r)=\frac{1}{\sqrt{-g_{tt}}}-1=\frac{1}{\sqrt{e^{A_{1}(\frac{r}{R})^{2}+a_{1}}}}-1.
\end{equation}
Utilizing Eq.\eqref{cc-1}, it can be expressed as
\begin{equation*}
Z_{s}(r)=\frac{1}{\sqrt{1-u}}-1.
\end{equation*}
We plot $Z_{s}(r)$ for the PS considering various values of $\zeta$.
Ivanov \cite{10-i} determined that the value of the anisotropic
configuration is 5.211. Figure \textbf{4} shows that $Z_{s}(r)$
remains positive and finite throughout the star's interior,
decreasing monotonically \cite{10-j,10-j1}. Furthermore, this
function stays within the specified limit $(Z_{s} < 5.211)$.
\begin{figure}\center
\epsfig{file=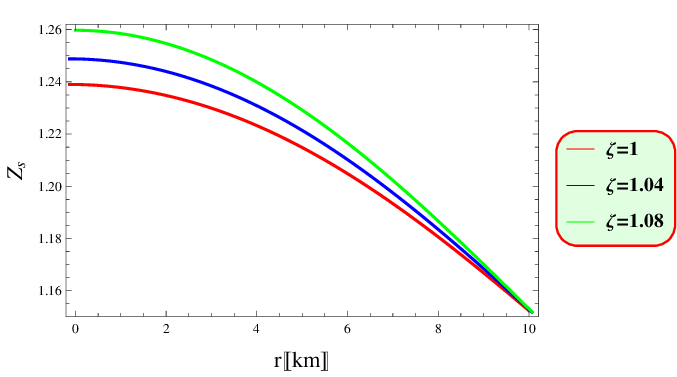,width=.55\linewidth}\caption{Plot of redshift
versus $r$.}
\end{figure}

\subsection{The Zeldovich Condition}

The Zeldovich condition is a fundamental criterion in astrophysics
that ensures the physical stability of matter under extreme
conditions, such as those found in neutron stars and other compact
objects. It states that the speed of sound within a material must
not exceed the speed of light, ensuring that the EoS remains causal.
This condition is derived from relativistic principles, emphasizing
that information or perturbations cannot propagate faster than the
speed of light. By imposing this restriction, the Zeldovich
condition plays a crucial role in determining the viable EoS for
dense matter, thereby constraining the physical and theoretical
models of cosmic structures. The Zeldovich condition is
mathematically expressed as follows
\begin{equation}\label{10}
\frac{P(0)}{\rho(0)}\leq 1,
\end{equation}
where $P(0)$ is the pressure at the center of the star and $\rho(0)$
is the central density. Let's confirm that this ratio does not
exceed 1. The values of $\rho$, $P_r$ and $P_{t}$ as $r\rightarrow
0$ can be determined as follows
\begin{align}\nonumber
\rho(0)&=\frac{\zeta (-2 a \eta - 17 \eta A_{1} - 18 A_{1})}{R^2 (-8
\eta^2 - 15 \eta - 6)},
\\\nonumber
P_{r}(0)&=\bigg[\frac{\zeta \big( 10a\eta^2 + 26a\eta + 12a -
11\eta^2 A_{1} - 13\eta A_{1} - 6A_{1} \big)}{R^2 (1 + \eta)
(-8\eta^2 - 15\eta - 6)} \bigg],
\\\nonumber
P_{t}(0)&=\frac{\zeta \big(-14a\eta^2 - 28a\eta - 12a + \eta^2 A_{1}
+ 5\eta A_{1} + 6 A_{1}\big)}{R^2 \big(-8\eta^3 - 23\eta^2 - 21\eta
- 6\big)}.
\end{align}
Utilizing the previously mentioned numerical values for the PS, we
can evaluate the Zeldovich inequality \eqref{10}. From the above
expressions, we can find $\frac{P_r(0)}{\rho(0)} = -0.0842$, which
is less than 1. Similarly, for $\frac{P_t(0)}{\rho(0)} = 0.3404$,
also less than 1. This confirms the validation of the Zeldovich
condition.

\subsection{Energy Conditions}

Astrophysical objects consist of a diverse range of materials and it
is crucial to distinguish between the types of matter (ordinary or
exotic) that make up these celestial structures. Energy constraints
are essential for assessing the possible configurations of fluid
matter within such systems. These constraints serve a vital role in
understanding specific cosmic structures and the interactions
between matter and energy under gravitational forces. For pulsar
stars, these constraints ensure the physical feasibility of their
internal matter configurations. The energy conditions are classified
into four main types.
\begin{itemize}
\item Null Energy Condition (NEC): This condition states that the energy density measured
along any null vector must be non-negative. Mathematically, it is
expressed as
\begin{equation*}
0 \leq \rho + P_r, \quad 0 \leq \rho + P_t.
\end{equation*}
This condition does not impose strict relations between matter
variables, allowing for scenarios where the pressure may exceed or
fall below the energy density.
\item Strong Energy Condition: This condition asserts that the sum of three times the
energy density and the pressure in any direction must be
non-negative. It is defined as
\begin{equation*}0 \leq \rho + P_r, \quad 0 \leq \rho + P_t, \quad 0 \leq \rho +
P_r + 2P_t.
\end{equation*}
This condition reflects the combined influence of energy density and
pressure in determining matter behavior under gravitational effects.
\item Dominant Energy Condition: According to this condition, the energy density must
remain non-negative and the corresponding vector field must be
time-like or null-like. Mathematically, it is represented as
\begin{equation*}
0 \leq \rho \pm P_r, \quad 0 \leq \rho \pm P_t.
\end{equation*}
This condition ensures that energy flows in physically consistent
ways across the system.
\item Weak Energy Condition: This condition requires that the energy density is
non-negative as measured by any observer. It is given by
\begin{equation*}
0 \leq \rho + P_r, \quad 0 \leq \rho + P_t, \quad 0 \leq \rho.
\end{equation*}
This condition is more stringent than the NEC because it applies to
all observers, ensuring the non-negativity of energy density in
every frame of reference.
\end{itemize}
These energy bounds substantially affect the existence of stable astrophysical
objects within spacetime. These conditions must be adhered by any feasible cosmic
structure. Figure \textbf{5} illustrates the energy conditions for different values
of $\zeta$, represented by the total stress-energy tensor. These plots confirm that
the proposed PS satisfies all energy conditions within the context of $f(\mathcal{Q},
\mathbb{T})$ gravity.
\begin{figure}
\epsfig{file=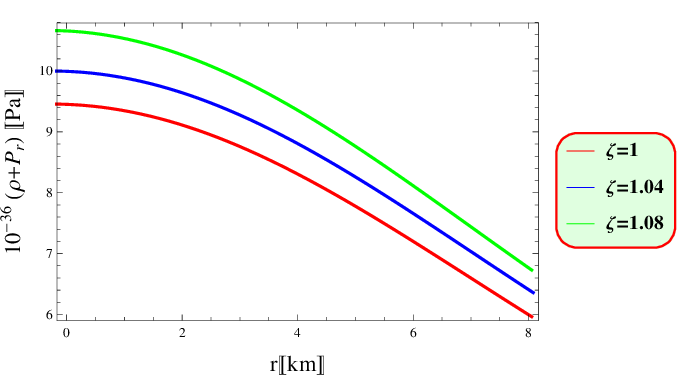,width=.52\linewidth} \epsfig{file=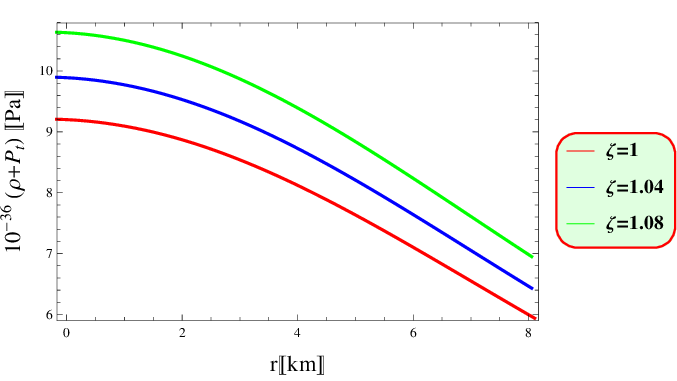,width=.52\linewidth}
\epsfig{file=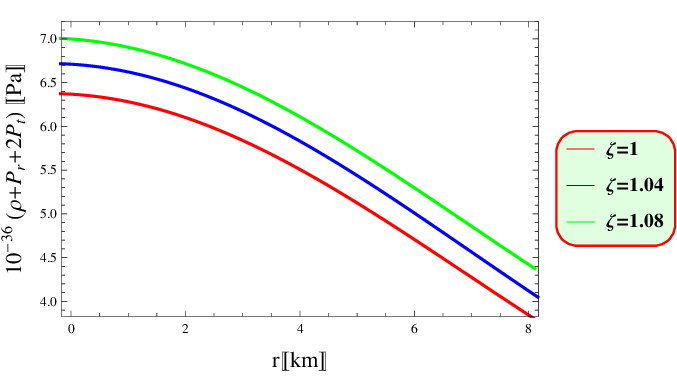,width=.52\linewidth} \epsfig{file=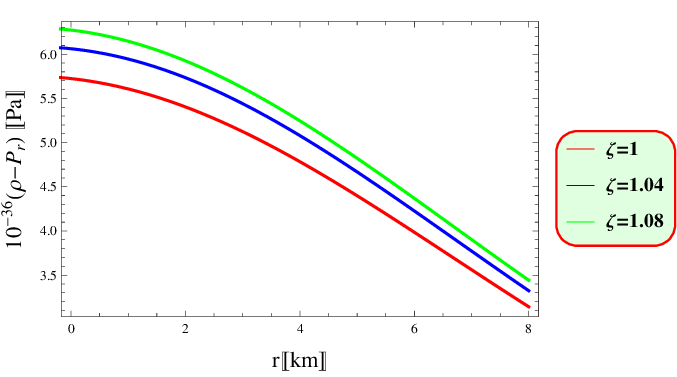,width=.52\linewidth}
\epsfig{file=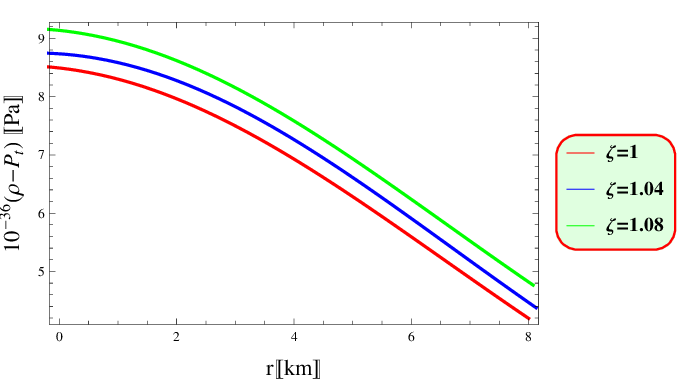,width=.52\linewidth} \caption{Plots of energy conditions versus
$r$.}
\end{figure}

\subsection{Causality Conditions}
\begin{figure}
\epsfig{file=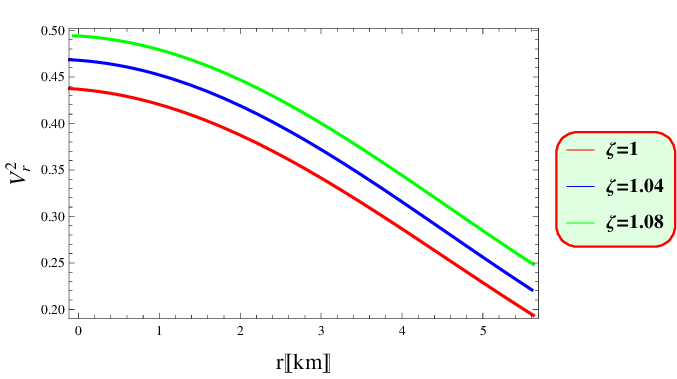,width=.52\linewidth} \epsfig{file=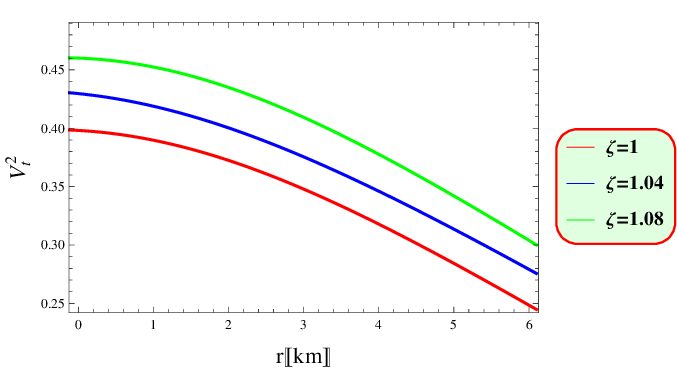,width=.52\linewidth}
\epsfig{file=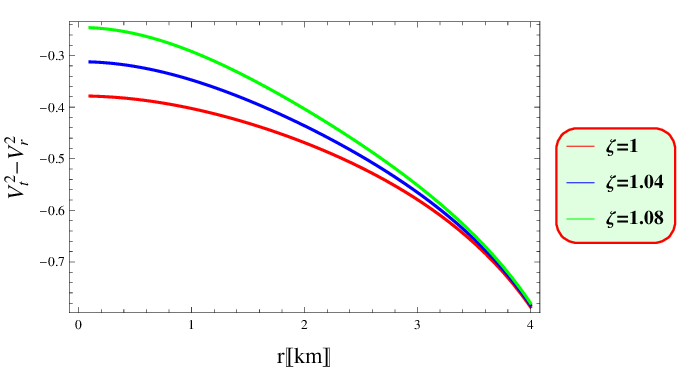,width=.52\linewidth} \caption{Plots of causality conditions
versus $r$.}
\end{figure}

The causality condition is a fundamental principle in physics that
ensures the relationship between cause and effect remains consistent
with the laws of relativity. It states that the speed of sound
represents the propagation of signals or disturbances within a
medium. This condition preserves the causal structure of spacetime,
preventing any violation of relativistic principles where
information or influence could travel faster than light. In
astrophysics, the causality condition plays a vital role in modeling
dense matter, such as in pulsar stars, by constraining equations of
state to physically realistic and stable configurations, ensuring
that the behavior of matter remains consistent with relativistic
causality. The causality condition asserts that the time-like
interval between any two events in spacetime must be zero or
positive, ensuring that no signal can exceed the speed of light.
Based on this, for an anisotropic fluid, this condition results in
\begin{equation}\label{11}
v_{r}^{2}= \frac{dP_r}{d\rho},\quad v_{t}^{2}= \frac{dP_t}{d\rho}.
\end{equation}
In a PS, the squared speed of sound must fall within the range $[0,
1]$ to maintain structural stability \cite{10-l}. Both radial and
tangential sound speeds meet the stability criteria, with $ 0 \leq
v_r^2 \leq 1 $ and $ 0 \leq v_t^2 \leq 1 $. The stability of a PS
can be evaluated by examining the cracking condition $0\leq \mid
v_t^2 - v_r^2 \mid \leq 1$ \cite{10-m}. Meeting this criterion
indicates that cosmic structures are stable and capable of
sustaining their configurations over time. Failure to satisfy this
condition suggests instability, potentially leading to the collapse
of the structure. This method provides researchers with a valuable
tool to assess the stability of cosmic formations, offering critical
insights into their behavior and role in the universe. The
expressions of the sound speed components are provided in Appendix
\textbf{A}. Figure \textbf{6} shows the radial and tangential
variations of the sound speed with respect to the radial distance.
We observe that $0.44 < v_r^2 < 0.50$ and $0.40 < v_t^2 < 0.46$.
Additionally, the condition $-0.79 < v_t^2 - v_r^2 < -0.37$ is
maintained throughout the interior of the PS, thereby ensuring
stability in its anisotropic stellar configurations.

\subsection{The Adiabatic Index and the Equilibrium of Hydrodynamic Forces}

To determine the stability of PS, we use the adiabatic index,
$\Gamma$, defined as \cite{10-p,10-p1}
\begin{equation*}
\Gamma=\frac{4}{3}\bigg(\frac{\sigma}{r|
\acute{P}_{r}|}+1\bigg)_{max},
\end{equation*}
\begin{equation*}
\Gamma_{r}=\frac{\rho+P_{r}}{P_{r}}v^{2}_{r},~~\Gamma_{t}
=\frac{\rho+P_{t}}{P_{t}}v^{2}_{t},
\end{equation*}
where the symbols $\Gamma_{r}$ and $\Gamma_{t}$ represent the radial
and tangential components of the adiabatic index, respectively.
Clearly, in the case of isotropy $(\sigma = 0)$, we get
$\frac{4}{3}$. For slightly anisotropic condition $(\sigma \neq 0)$,
similar to those in Newtonian theory, $\sigma$ is less than
$\frac{4}{3}$, which is consistent with the usual stability
criterion \cite{10-q}. The adiabatic index can be found in Appendix
\textbf{B}. Figure \textbf{7} shows a stable anisotropic
distribution for the PS across various $\zeta$ values.
\begin{figure}
\epsfig{file=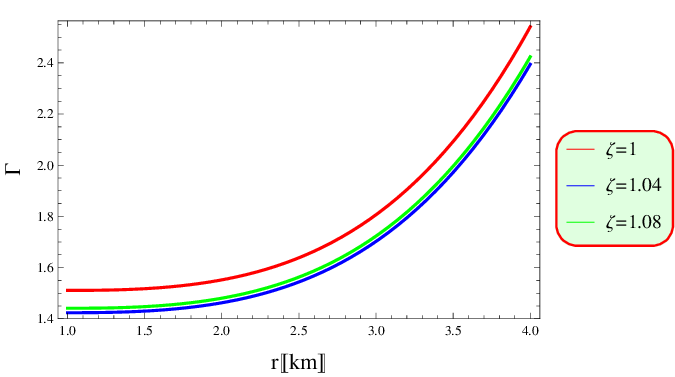,width=.52\linewidth} \epsfig{file=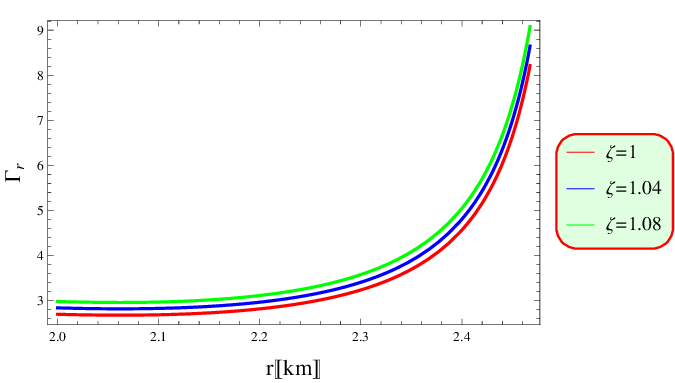,width=.52\linewidth}
\epsfig{file=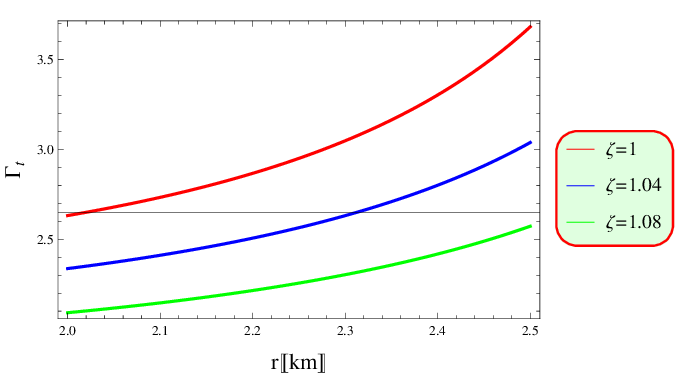,width=.52\linewidth} \caption{Plots of adiabatic index against
$r$.}
\end{figure}

The TOV equation is a key formula in astrophysics that describes the
equilibrium of a static spherically symmetric star. It helps to
explain how the pressure within a star balances the gravitational
forces to maintain stability. It is also essential for understanding
the structure and behavior of PS and other compact objects. By
solving this equation, scientists can predict the internal pressure,
density and overall stability of these dense celestial bodies. The
TOV equation for an anisotropic matter distribution is formulated as
follows \cite{10-r}
\begin{equation}\label{d}
M_{G}(r) e^{\frac{\alpha-\beta}{2}} \frac{1}{r^2} (\rho + P_r) +
\frac{dP_r}{dr}-\frac{2}{r} (P_t - P_r) = 0,
\end{equation}
where the gravitational mass, denoted as $M_{G}$, is defined as
\cite{10-s,10-s1}
\begin{equation}\label{d1}
M_{G}(r) = 4\pi \int (\mathbb{T}^t_t- \mathbb{T}^r_r -
\mathbb{T}^\phi_\phi- \mathbb{T}^\theta_\theta) r^2
e^{\frac{\alpha+\beta}{2}} dr.
\end{equation}
Solving this equation, we obtain
\begin{equation}\label{d2}
M_{G}(r) = \frac{1}{2} r^2 e^{\frac{\beta-\alpha}{2}}
\alpha^{\prime}.
\end{equation}
Substituting Eq.\eqref{d2} into \eqref{d}, we have
\begin{equation}\label{33}
\frac{1}{2} \alpha^{\prime} (\rho + P_r) + \frac{dP_r}{dr} -
\frac{2}{r} (P_{t} - P_r) = 0.
\end{equation}
We analyze the hydrodynamic equilibrium of our model by utilizing
the TOV equation modified for $f(\mathcal{Q}, \mathbb{T})$ gravity
\begin{equation}
F_{a}+F_{g}+F_{h}+F_{(\mathcal{Q}, \mathbb{T})}=0,
\end{equation}
here, $F_{a}$, $F_{g}$ and $F_{h}$ denote the anisotropic,
gravitational, hydrostatic forces, respectively, along with the
force $F_{(\mathcal{Q}, \mathbb{T})}$. These forces are described as
follows
\begin{eqnarray}\nonumber
F_{a} &=& 2\frac{\sigma}{r},\quad F_{g}= \frac{\alpha^{\prime}
(\rho+P_{r})}{2},\\
\nonumber F_{h}&=& -P_{r}^{\prime},\quad
F_{(\mathcal{Q},\mathbb{T})}=
P_{r}^{\prime}+\frac{\beta^{\prime}(r)}{2}(\rho-P_{r})-\frac{2}{r}(P_{t}-P_{r}).
\end{eqnarray}
Simplifying the above values, we obtain
\begin{eqnarray}\nonumber
F_{a}&=&\bigg[6 \zeta  e^{-\frac{A_{1} r^2}{R^2}} \big(a^2 r^4
\big(8 \eta ^2 r^2 e^{\frac{A_{1} r^2}{R^2}}+\eta  \big(6 r^2
e^{\frac{A_{1} r^2}{R^2}}-3\big)-2\big)a r^2 \big(A_{1} r^2 \big(8
\eta ^2 r^2
\\\nonumber
&-& e^{\frac{A_{1} r^2}{R^2}}+\eta \big(6 r^2 e^{\frac{A_{1}
r^2}{R^2}}-3\big)-2\big)+2 R^2 \big(2 \big(2 \eta ^2+3 \eta +1\big)
r^2 e^{\frac{2 A_{1} r^2}{R^2}}-\big(2 \eta ^2
\\\nonumber
&+&3 \eta+2 \big(6 \eta ^2+6 \eta +1\big) r^2+1\big) e^{\frac{A_{1}
r^2}{R^2}}+6 \eta ^2+12 \eta +5\big)\big)-R^2 \big(4 f \eta ^2 r^2
\\\nonumber
&+&5 A_{1} \eta  r^2+2 r^2 e^{\frac{2 A_{1} r^2}{R^2}} \big(2 f
\big(2 \eta ^2+3 \eta +1\big) r^2+\eta R^2\big)-e^{\frac{ r^2}{R^2}}
\big(2 A_{1} r^2 \big(2 \eta ^2
\\\nonumber
&+&+3 \eta(\eta +2) r^2+1\big)+R^2 \big(4 \eta ^2+\eta \big(2
r^2+5\big)+2\big)\big)+4 \eta ^2 R^2+5 \eta  R^2
\\\nonumber
&+&2 R^2\big)\big)\bigg]\bigg[(\eta +1) r^3 R^4 \big(8 \eta ^2
\big(r^2 e^{\frac{A_{1} r^2}{R^2}}-1\big)+3 \eta \big(2 r^2
e^{\frac{A_{1} r^2}{R^2}}-5\big)-6\big)\bigg]^{-1},
\end{eqnarray}
\begin{eqnarray}\nonumber
F_{g}&=&-\bigg[a \zeta  e^{-\frac{A_{1} r^2}{R^2}} \big(a^2 \eta  (4
\eta +3) r^4 \big(2 r^2 e^{\frac{A_{1} r^2}{R^2}}+1\big)-a r^2
\big(R^2 \big(6 \big(2 \eta ^2+3 \eta +1\big)
\\\nonumber
&\times& r^2 e^{\frac{2 A_{1} r^2}{R^2}}-\big(6 \eta ^2+9 \eta
+\big(28 \eta ^2+30 \eta +6\big) r^2+3\big) e^{\frac{A_{1}
r^2}{R^2}}+10 \eta ^2+21 \eta
\\\nonumber
&+&9\big)A_{1} \eta (4 \eta +3) r^2 \big(2 r^2 e^{\frac{A_{1}
r^2}{R^2}}+1\big)\big)+R^2 \big(-A_{1} r^2 \big(6 \big(2 \eta ^2+3
\eta +1\big) r^2 e^{\frac{2 A_{1} r^2}{R^2}}
\\\nonumber
&+&\big(\big(4 \eta ^2-6\big) r^2-3 \big(2 \eta ^2+3 \eta
+1\big)\big) e^{\frac{A_{1} r^2}{R^2}}+2 \eta ^2-3\big)-R^2
\big(e^{\frac{A_{1} r^2}{R^2}}-1\big) \big(2 \eta ^2
\\\nonumber
&\times&\big(2 r^2 e^{\frac{A_{1} r^2}{R^2}}-5\big)+3 \eta  \big(2
r^2 e^{\frac{A_{1}
r^2}{R^2}}-5\big)-6\big)\big)\big)\bigg]\bigg[(\eta +1) r R^6 \big(8
\eta ^2 \big(r^2 e^{\frac{A_{1} r^2}{R^2}}-1\big)
\\\nonumber
&+&3 \eta \big(2 r^2 e^{\frac{A_{1}
r^2}{R^2}}-5\big)-6\big)\bigg]^{-1},
\end{eqnarray}
\begin{eqnarray}\nonumber
F_{h}&=&\bigg[2 e^{-\frac{A_{1} r^2}{R^2}} \zeta  \big(a^2
\big(A_{1} r^2 \big(16 \big(4 e^{\frac{2 A_{1} r^2}{R^2}} r^4-6
e^{\frac{A_{1} r^2}{R^2}} r^2+3\big) \eta ^4+2 \big(48 e^{\frac{2
A_{1} r^2}{R^2}} r^4
\\\nonumber
&-&140e^{\frac{A_{1} r^2}{R^2}} r^2+97\big)\eta ^3+9 \big(4
e^{\frac{2 A_{1}r^2}{R^2}} r^4-28 e^{\frac{A_{1} r^2}{R^2}}
r^2+31\big) \eta ^2-24 \big(3 e^{\frac{A_{1} r^2}{R^2}}
\\\nonumber
&\times&r^2-7\big) \eta +36\big)+36\big)-R^2 \big(16 \big(4
e^{\frac{2 A_{1} r^2}{R^2}} r^4-8 e^{\frac{A_{1} r^2}{R^2}}
r^2+3\big) \eta ^4
\\\nonumber
&+&2 \big(48 e^{\frac{2 A_{1} r^2}{R^2}} r^4-168 e^{\frac{A_{1}
r^2}{R^2}} r^2+97\big) \eta ^3+3 \big(12 e^{\frac{2 A_{1} r^2}{R^2}}
r^4-92 e^{\frac{A_{1} r^2}{R^2}} r^2+93\big) \eta ^2
\\\nonumber
&-&24 \big(3 e^{\frac{A_{1} r^2}{R^2}} r^2-7\big) \eta +36\big)\big)
r^4-a \big(4 e^{\frac{A_{1} r^2}{R^2}} \eta (\eta +1)
\big(e^{\frac{A_{1} r^2}{R^2}} \big(2 \eta ^2+6 \eta +3\big)
\\\nonumber
&-&3 \big(2 \eta ^2+3 \eta +1\big)\big) R^4-A_{1} \big(8 \big(-46
e^{\frac{A_{1} r^2}{R^2}} r^2+e^{\frac{2 A_{1} r^2}{R^2}} \big(26
r^2-1\big) r^2+21\big) \eta ^4
\\\nonumber
&+&\big(-980 e^{\frac{A_{1} r^2}{R^2}} r^2+4 e^{\frac{2 A_{1}
r^2}{R^2}} \big(79 r^2-8\big) r^2+651\big) \eta ^3+12 \big(-68
e^{\frac{A_{1} r^2}{R^2}} r^2
\\\nonumber
&+&e^{\frac{2 A_{1} r^2}{R^2}} \big(10 r^2-3\big) r^2+75\big) \eta
^2-6 \big(36 e^{\frac{A_{1} r^2}{R^2}} r^2+2 e^{\frac{2
A_{1}r^2}{R^2}} r^2-87\big) \eta +108\big) R^2
\\\nonumber
&+&A_{1}^2 r^2 \big(16 \big(4 e^{\frac{2 A_{1} r^2}{R^2}} r^4-6
e^{\frac{A_{1} r^2}{R^2}} r^2+3\big) \eta ^4+2 \big(48 e^{\frac{2
A_{1} r^2}{R^2}} r^4-140 e^{\frac{A_{1} r^2}{R^2}} r^2+97\big) \eta
^3
\\\nonumber
&+&9 \big(4 e^{\frac{2 A_{1} r^2}{R^2}} r^4-28 e^{\frac{A_{1}
r^2}{R^2}} r^2+31\big) \eta ^2-24 \big(3 e^{\frac{A_{1} r^2}{R^2}}
r^2-7\big) \eta +36\big)\big) r^4+R^2
\\\nonumber
&\times&\big(-A_{1}^2\big(8 \big(-2 e^{\frac{A_{1}
r^2}{R^2}}r^2+e^{\frac{2 A_{1}r^2}{R^2}} \big(2 r^4+r^2\big)+1\big)
\eta ^4+\big(-108 e^{\frac{A_{1} r^2}{R^2}} r^2
\\\nonumber
&+&4 e^{\frac{2 A_{1} r^2}{R^2}} \big(11 r^2+8\big) r^2+63\big) \eta
^3+12 \big(-14 e^{\frac{A_{1} r^2}{R^2}} r^2+e^{\frac{2 A_{1}
r^2}{R^2}} \big(2 r^2+3\big) r^2
\\\nonumber
&+&12\big) \eta ^2+6 \big(-12 e^{\frac{A_{1} r^2}{R^2}} r^2+2
e^{\frac{2 A_{1} r^2}{R^2}} r^2+21\big) \eta +36\big) r^4-A_{1} R^2
\eta \big(8 \eta ^3+31 \eta ^2
\\\nonumber
&+&36 \eta -4 e^{\frac{A_{1} r^2}{R^2}} r^2 \big(2 \eta ^3+9 \eta
^2+9 \eta +3\big)+4 e^{\frac{2A_{1}r^2}{R^2}} r^2 \big(4 r^2 \eta
^33 \big(r^2+2\big) \eta ^2+9 \eta
\\\nonumber
&+&3\big)+12\big) r^2+\big(-1+e^{\frac{A_{1} r^2}{R^2}}\big) R^4
\eta \big(8 \big(2 e^{\frac{2 A_{1} r^2}{R^2}} r^4-2 e^{\frac{A_{1}
r^2}{R^2}} r^2+1\big) \eta ^3+\big(12
\\\nonumber
&\times&e^{\frac{2 A_{1} r^2}{R^2}} r^4-44 e^{\frac{A_{1} r^2}{R^2}}
r^2+31\big) \eta ^2+\big(36-24 e^{\frac{A_{1} r^2}{R^2}} r^2\big)
\eta +12\big)\big)\big)\bigg]\bigg[r^3 R^6 (\eta +1)
\\\nonumber
&\times& \big(8 \big(e^{\frac{A_{1} r^2}{R^2}} r^2-1\big) \eta ^2+3
\big(2 e^{\frac{A_{1} r^2}{R^2}} r^2-5\big) \eta
-6\big)^2\bigg]^{-1}.
\end{eqnarray}
Figure \textbf{8} shows that our PS remains in equilibrium. This
equilibrium is attained because the cumulative effects of $F_{a}$,
$F_{g}$, $F_{h}$, and $F_{(\mathcal{Q}, \mathbb{T})}$ result in a
net sum of zero. This results in a stable model for PS across
different values of $\zeta$.
\begin{figure}
\epsfig{file=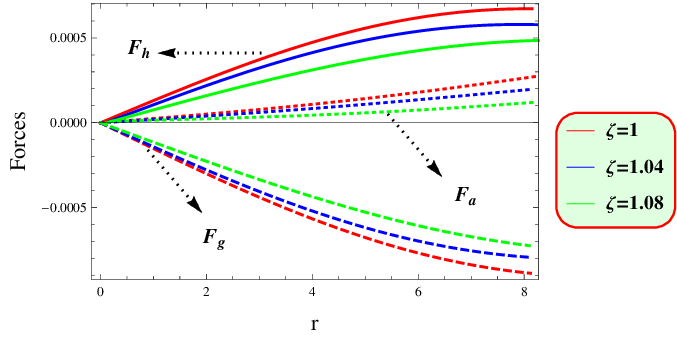,width=.52\linewidth}
\epsfig{file=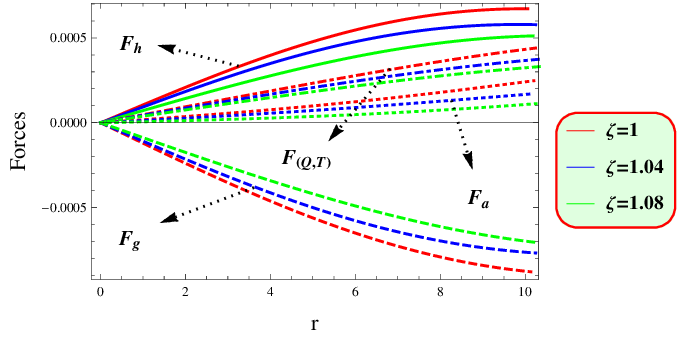,width=.52\linewidth}\caption{Plots of the TOV equation versus
$r$.}
\end{figure}

\section{Equation of State Parameter and Compactness}

Here, we utilize the following relationships \cite{7-e}
\begin{align}\label{E}
P_{r}(\rho) \approx v^2_{r} (\rho - \rho_I), \quad P_{t}(\rho)
\approx v^{2}_{t} (\rho - \rho_{II}),
\end{align}
where $\rho_{I}$ refers to the density at which the radial pressure
$P_{r(\rho_I)} = 0$, $\rho_I$ and $\rho_{II}$ denote the densities
at the star's surface associated with the radial and tangential
pressures, respectively. It is important to note that, unlike
$\rho_{I}$ which results in $P_{r}$ being zero, $\rho_{II}$ does not
necessarily cause $P_{t}$ to be zero. The current model fully
determines the speed of sound and surface density. For example, when
$\zeta = 1.08$, using Eqs.\eqref{38} and \eqref{39}, we obtain the
following values $v_{r}^{2} \approx 0.49, v_{t}^{2} \approx 0.46$,
$\rho_{I} = 5.4 \times 10^{14}$ g/cm$^{3}$, and $\rho_{II} \approx
4.3 \times 10^{14}$ g/cm$^{3}$. For $\zeta =1.04$, $v_{r}^{2}
\approx 0.47, v_{t}^{2} \approx 0.43$, $\rho_{I} = 5.6 \times
10^{14}$ g/cm$^{3}$, and $\rho_{II} \approx 4.2 \times 10^{14}$
g/cm$^{3}$. When $\zeta =1$, $v_{r}^{2} \approx 0.44, v_{t}^{2}
\approx 0.39$, $\rho_{I} = 5.9 \times 10^{14}$ g/cm$^{3}$, and
$\rho_{II} \approx 4 \times 10^{14}$ g/cm$^{3}$.

The EoS parameter is crucial for characterizing the relationship
between $P$ and $\rho$ across various physical systems. For a model
to be physically viable, the radial and tangential EoS parameters
must lie within the range $[0,1]$ \cite{10-n}. This is given as
\begin{equation}\label{C}
\omega_{r}= \frac{P_{r}}{\rho},\quad \omega_{t}= \frac{P_{t}}{\rho}.
\end{equation}
Inserting Eqs.\eqref{5a}-\eqref{5c} into \eqref{C}, we obtain
\begin{align}\nonumber
\omega_{r}&=-\bigg[R^4 \big(\frac{1}{R^4}\bigg\{2 r^2
\big(-e^{\frac{A_{1} r^2}{R^2}} \big(4 a^2 \eta  (4 \eta +3) r^4+a
\big(R^2 \big((\eta +1) (11 \eta +6)+A_{1} R^2
\\\nonumber
&\times\big((\eta +1) (11 \eta +6)+2 (\eta (27 \eta +29)+6)
r^2\big)-4 A_{1} \eta  (4 \eta +3) r^4\big)+2 (\eta +2)
\\\nonumber
&\times(\eta +3) r^2\big)\big) -6 a^2 \eta ^2 r^2-4 a^2 \eta r^2+6 a
f \eta ^2 r^2+4 a A_{1} \eta r^2+2(\eta +1)(11 \eta +6)
\\\nonumber
&\times r^2 R^2 (a+A_{1}) e^{\frac{2 A_{1} r^2}{R^2}}+21 a \eta ^2
R^2 +43 a \eta R^2+18 a R^2+13 f \eta ^2 R^2+21 A_{1} \eta R^2
\\\nonumber
&+6 A_{1} R^2\big)\bigg\}+2 \big(e^{\frac{A_{1} r^2}{R^2}}-1\big)
\big(\eta \big(2 (\eta +3) r^2 e^{\frac{A_{1} r^2}{R^2}}-13 \eta
-17\big)-6\big)\big)\bigg]\bigg[2 (\eta +1)
\\\nonumber
&\times\big(2 a^2 \eta r^4-2 a A_{1} \eta r^4-2 \eta  r^2 R^2
e^{\frac{2 A_{1} r^2}{R^2}} \big(r^2 (a+A_{1})+3 R^2\big)+R^2
e^{\frac{A_{1} r^2}{R^2}}\big(\eta  r^2 \big(2 r^2
\\\nonumber
&\times(a-5 A_{1}) +a+A_{1}\big)+R^2 \big(\eta \big(6
r^2+7\big)+6\big)\big) +a \eta r^2 R^2 +3 R^2\big)+9 A_{1} \eta  r^2
R^2
\\\nonumber
&+12 A_{1} r^2 R^2-7 \eta R^4-6 R^4\big)\bigg]^{-1},
\\\nonumber
\omega_{t}&=-\bigg[\big(\eta  \big(2 (4 \eta +3) r^2 e^{\frac{A_{1}
r^2}{R^2}}-8 \eta -15\big)-6\big) \big(a^2 r^4 \big(\eta  \big(2 (4
\eta +3) r^2 e^{\frac{A_{1} r^2}{R^2}}-6 \eta -13\big)
\\\nonumber
&-6\big)+a r^2 \big(A_{1} r^2 \big(\eta  \big(-2 (4 \eta +3) r^2
e^{\frac{A_{1} r^2}{R^2}}+6 \eta +13\big)+6\big)+R^2 \big(\eta
\big(e^{\frac{A_{1} r^2}{R^2}}\big(2 r^2
\\\nonumber
&\times \big(-(\eta +1) e^{\frac{A_{1} r^2}{R^2}}+9 \eta
+7\big)+\eta +1\big)-15 \eta -29\big)-12\big)\big)+R^2 \big(A_{1}
r^2 \big(\eta
\\\nonumber
&\times\big(e^{\frac{A_{1}r^2}{R^2}} \big(-2 r^2 \big((\eta +1)
e^{\frac{A_{1} r^2}{R^2}}+\eta +2\big)+\eta +1\big)+\eta
+6\big)+6\big) +\eta R^2 \big(e^{\frac{A_{1} r^2}{R^2}}
\\\nonumber
&-1\big)\big(\eta \big(2 r^2 e^{\frac{A_{1}
r^2}{R^2}}-1\big)-2\big)\big)\big)\bigg] \bigg[\big(\eta \big(2
(\eta +1) (4 \eta +3) r^2 e^{\frac{A_{1} r^2}{R^2}}-\eta  (8 \eta
+23)
\\\nonumber
&-21\big)-6\big )\big(2 a^2 \eta  r^4-2 a A_{1} \eta  r^4-2 \eta r^2
R^2 e^{\frac{2 A_{1} r^2}{R^2}} \big(r^2 (a+A_{1}) +3 R^2\big)+R^2
\\\nonumber
&\times e^{\frac{A_{1} r^2}{R^2}} \big(\eta  r^2 \big(2 r^2(a-5
A_{1})+a+A_{1}\big)+R^2 \big(\eta \big(6 r^2+7\big)+6\big)\big)+a
\eta r^2 R^2
\\\nonumber
&+9 A_{1} \eta  r^2 R^2+12 A_{1} r^2 R^2-7 \eta  R^4-6
R^4\big)\bigg]^{-1}.
\end{align}
\begin{figure}
\epsfig{file=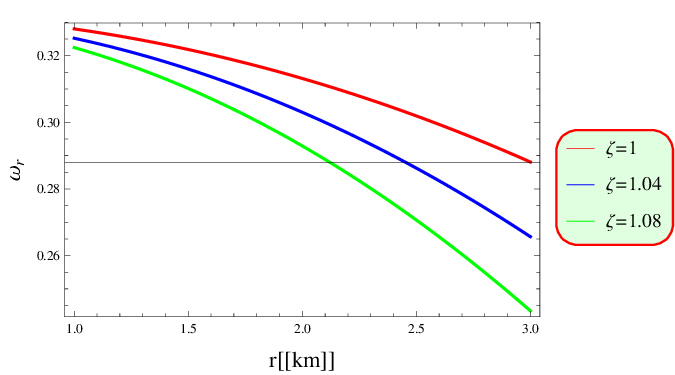,width=.52\linewidth} \epsfig{file=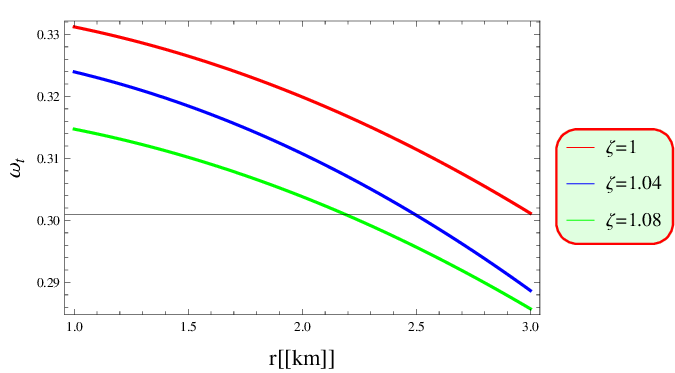,width=.52\linewidth}
\epsfig{file=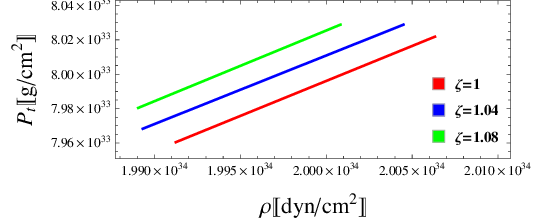,width=.52\linewidth} \epsfig{file=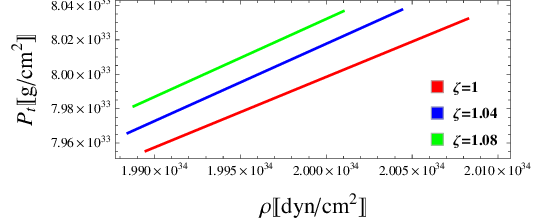,width=.52\linewidth}
\caption{Plots of EoS versus $r$.}
\end{figure}

Figure \textbf{9} illustrates the best-fit EoS for the PS, depicting
the graphical behavior of density and radial pressure across various
values of $\zeta$. These results align well with a linear EoS
pattern. They also meet the viability criteria for the PS.
Similarly, the tangential EoS exhibits a strong correlation with a
linear model. The derived EoSs are primarily valid near the star's
center and throughout its interior within the range of $[0,1]$
\cite{10-o}. The compactness function $u = \frac{M(r)}{r}$ is
essential in assessing the viability of a PS. Buchdahl \cite{10-h}
proposed a limit for the mass-radius ratio, stating that $u$ must be
less than $\frac{4}{9}$ for a PS to be viable. Figure \textbf{10}
demonstrates that compactness increases steadily remaining within
the specified limit of $(u < \frac{4}{9})$.
\begin{figure}\center
\epsfig{file=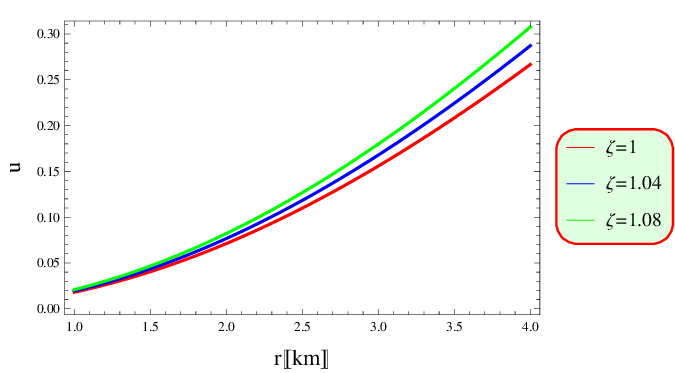,width=.55\linewidth} \caption{Plot of compactness versus $r$.}
\end{figure}

\section{Concluding Remarks}

In this paper, we have explored PS in the framework of
$f(\mathcal{Q}, \mathbb{T})$ theory of gravity for a particular
model $f(\mathcal{Q}, \mathbb{T}) = \zeta \mathcal{Q} + \eta
\mathbb{T}$. We have explored how this theory changes our
understanding of PS and found that these modifications significantly
affect its composition and dynamics. Focusing on an anisotropic
fluid scenario and assuming the KB ansatz for the star's inner
region, we have identified crucial limits on our model parameters.
The key results are summarized as follows.
\begin{itemize}
\item
We have shown that $\rho $, $P_{r}$ and $P_{t}$ for various values
of $\zeta$ are highest at the center and decrease towards the
surface. The radial pressure $P_{r}$ reaches zero at the surface of
the star (Figure \textbf{1}).
\item
The anisotropy is zero at the center where $P_{t} = P_{r}$ but
becomes positive in other regions (Figure \textbf{2}). This behavior
indicates that the model aligns with theoretical predictions and
validates the physical viability of the proposed stellar structure.
\item
The mass function for the PS clearly demonstrates that the mass of
the star steadily and uniformly increases with the radius (Figure
\textbf{3}). This observation aligns with the stability of the PS.
\item
The redshift function decreases monotonically and remains within the
$Z_{s} < 5.211$ limit (Figure \textbf{4}). This is consistent with
theoretical expectations.
\item
The energy conditions for various values of $\zeta$ confirm that the
proposed PS meets all energy conditions, validating the physical
viability of our model (Figure \textbf{5}).
\item
We have observed that the sound speed propagates in both radial and tangential
directions for PS at various values of $\zeta$ (Figure \textbf{6}). Furthermore, we
have found that $(-1 < |v_t^2 - v_r^2| < 0)$ throughout the pulsar's interior,
ensuring stability in its anisotropic stellar configurations.
\item
The condition $\Gamma > \frac{4}{3}$ is satisfied (Figure
\textbf{7}), validating the physical stability of PS for various
values of $\zeta$.
\item
We have determined that equilibrium is achieved as the sum of the
combined forces $F_{a}$, $F_{g}$, $F_{h}$ and $F_{(\mathcal{Q},
\mathbb{T})}$ equals to zero (Figure \textbf{8}). This leads to a
stable model for the PS across various values of $\zeta$.
\item
The EoS for the PS demonstrates a strong linear relationship between
density and radial pressure for all values of $\zeta$. The
tangential pressure also fits well with a linear model. These
equations hold true near the center and throughout the interior of
the star within the range of 0 to 1. This confirms that $\omega_{r}$
and $\omega_{t}$ meet the criteria for PS viability, suggesting
enhanced stability (Figure \textbf{9}).
\item
The compactness gradually increases throughout the star's interior
and remains within the specified limits of $u < \frac{4}{9}$ for
various values of $\zeta$ (Figure \textbf{10}).
\end{itemize}

Our results indicate that $f(\mathcal{Q}, \mathbb{T})$ gravity
offers a new and fascinating perspective for understanding PS. It is
worth noting that this theory satisfies both the stability criteria
and empirical observations. Moreover, our findings align with the
recent literature \cite{7-e} that support the viability of
$f(\mathcal{Q}, \mathbb{T})$ gravity in explaining the complexities
of astrophysical phenomena. The continued exploration and
advancement in this domain hold significant potential for enhancing
our understanding of the universe. Future research in this area
promises to refine and rigorously test these models, potentially
revealing the new insights into the fundamental nature of gravity
and the dynamics of compact objects like PS.

Despite these advancements, the application of $f(\mathcal{Q},
\mathbb{T})$ gravity to strong-field regimes, particularly compact
astrophysical objects like neutron stars and pulsars, remains under
explored. Compact stars, formed during supernova collapses, exhibit
extreme gravitational and matter densities, making them ideal
laboratories for testing modifications to GR. The PS, with its
observed mass and radius properties, provides a compelling case for
exploring these theories in anisotropic stellar configurations.
Through our analysis, we have demonstrated that the PS within the
$f(\mathcal{Q}, \mathbb{T})$ modified gravity framework exhibits a
dense and stable profile. Key physical quantities, including
density, pressure and anisotropy, show extended ranges that make
this star physically viable and stable, even at its center. This
marks a significant improvement over predictions from GR and other
modified gravity theories, where compact stars often fail to
maintain stability in central regions. To validate our findings, we
have analyzed critical parameters such as energy conditions,
compactness, redshift and the adiabatic index, which further confirm
the dense and stable nature of the PS within this framework. By
systematically exploring the impact of $f(\mathcal{Q}, \mathbb{T})$
gravity, this study addresses existing gaps in the literature and
advances our understanding of strong-field astrophysics, offering a
more viable and comprehensive description of anisotropic stellar
structures.

The $f(\mathcal{Q}, \mathbb{T})$ gravity is a powerful extension of
GR. This theory broadens GR applicability to regions with strong
gravitational fields where GR often struggles. One major strength of
this theory is its ability to model anisotropic stellar
configurations, making it particularly suitable for compact objects
such as pulsars. In this study, stable configurations have been
successfully predicted for PS, satisfying essential physical
criteria like the mass-radius relation, redshift, causality
conditions and the Zeldovich condition. Furthermore, the theory
aligns well with observational data, providing a robust framework
for describing compact stellar structures while accommodating
complex matter distributions. By directly linking geometry with
matter, $f(\mathcal{Q}, \mathbb{T})$ gravity unifies gravitational
and matter interactions in a way that GR cannot achieve in
strong-field regimes.

Despite its strengths, $f(\mathcal{Q}, \mathbb{T})$ gravity has
certain limitations that require further investigation. The theory's
predictions depend on the parameters $\zeta$ and $\eta$, which
govern the functional form of the theory. These parameters currently
lack direct observational constraints and aligning theoretical
results with astrophysical data necessitates fine adjustments. This
sensitivity reduces the universality of the theory and limits its
predictive power. Moreover, the mathematical complexity of the field
equations poses another challenge, as deriving analytical solutions
is often difficult, leading to a reliance on numerical methods. Such
reliance on numerical approximations can introduce uncertainties,
particularly when modeling precise astrophysical phenomena.
Additionally, while this study has demonstrated the theory
applicability to a single pulsar, its application to a broader range
of compact objects and cosmological phenomena remains to be tested.

In summary, $f(\mathcal{Q}, \mathbb{T})$ gravity is a significant
and robust theoretical framework designed to explain compact stars
and observational data, particularly in strong-field gravitational
environments. Its ability to unify geometry and matter interactions
makes it a promising extension of GR. However, its dependence on
tunable parameters and the complexity of its equations highlight the
need for further research. By pursuing applications in astrophysics,
gravitational wave science and cosmology, and refining its
mathematical framework, $f(\mathcal{Q}, \mathbb{T})$ gravity has the
potential to play a pivotal role in advancing our understanding of
the universe.

\section*{Appendix A: Casuality Condition}
\renewcommand{\theequation}{A\arabic{equation}}
\setcounter{equation}{0} The expressions of squared sound speed in
the radial and tangential directions are given as follows.
\begin{eqnarray}\nonumber
v_{r}^{2}&=&\bigg[(\eta +1) \big(A_{1}^2 \big(2 a \eta  \big(8
\big(2 e^{\frac{A_{1} r^2}{R^2}} r^2-1\big) \eta ^2+3 \big(4
e^{\frac{A_{1} r^2}{R^2}} r^2-5\big) \eta -6\big) r^2
\\\nonumber
&+&R^2\big(8 \big(-18 e^{\frac{A_{1} r^2}{R^2}} r^2+e^{\frac{2 A_{1}
r^2}{R^2}} \big(10 r^4+r^2\big)+9\big) \eta ^3+3 \big(-100
e^{\frac{A_{1} r^2}{R^2}} r^2+4
\\\nonumber
&\times&e^{\frac{2 A_{1} r^2}{R^2}} \big(5 r^2+2\big) r^2+77\big)
\eta ^2+6 \big(-24 e^{\frac{A_{1} r^2}{R^2}} r^2+2 e^{\frac{2
A_{1}r^2}{R^2}} r^2+39\big) \eta
\\\nonumber
&+&72\big)\big) r^4+A_{1} \big(-2 a^2 \eta \big(8 \big(2
e^{\frac{A_{1} r^2}{R^2}} r^2-1\big) \eta ^2+3 \big(4 e^{\frac{A_{1}
r^2}{R^2}} r^2-5\big) \eta -6\big) r^4
\\\nonumber
&-&a R^2 \eta \big(4 e^{\frac{A_{1} r^2}{R^2}} \eta (4 \eta +3)
r^2+4 e^{\frac{2 A_{1} r^2}{R^2}} \big(\big(4 r^2-2\big) \eta ^2+3
\big(r^2-2\big) \eta -3\big) r^2
\\\nonumber
&-&3 \big(8 \eta ^2+15 \eta +6\big)\big) r^2+R^4 \big(-8 \big(6
e^{\frac{2 A_{1} r^2}{R^2}} r^4-15 e^{\frac{A_{1} r^2}{R^2}}
r^2+7\big) \eta ^3
\\\nonumber
&-&3 \big(-60 e^{\frac{a r^2}{R^2}} r^2+4 e^{\frac{2 A_{1} r^2}{R^2}} \big(3
r^2-2\big) r^2+51\big) \eta ^2+12 \big(5 e^{\frac{A_{1} r^2}{R^2}} r^2+e^{\frac{2
A_{1} r^2}{R^2}} r^2
\\\nonumber
&-&11\big) \eta -36\big)\big) r^2+R^2 \big(-2 a^2 \eta \big(8 \eta
^2+15 \eta +6\big) r^4+4 a e^{\frac{A_{1} r^2}{R^2}} R^2 \eta
\big(e^{\frac{A_{1} r^2}{R^2}}
\\\nonumber
&\times&\big(2 \eta ^2+6 \eta +3\big)-3 \big(2 \eta ^2+3 \eta
+1\big)\big) r^4+\big(-1+e^{\frac{A_{1} r^2}{R^2}}\big) R^4 \big(8
\big(6 e^{\frac{2 A_{1} r^2}{R^2}} r^4
\\\nonumber
&-&14 e^{\frac{A_{1} r^2}{R^2}} r^2+7\big) \eta ^3+9 \big(4
e^{\frac{2 A_{1} r^2}{R^2}} r^4-20 e^{\frac{A_{1} r^2}{R^2}}
r^2+17\big) \eta ^2+\big(132-72 e^{\frac{A_{1} r^2}{R^2}} r^2\big)
\eta
\\\nonumber
&+&36\big)\big)\big)\bigg]\bigg[-2 a^2 \eta \big(A_{1} r^2 \big(8
\big(8 e^{\frac{2 A_{1} r^2}{R^2}} r^4+6 e^{\frac{A_{1} r^2}{R^2}}
r^2-3\big) \eta ^3+\big(96 e^{\frac{2 A_{1} r^2}{R^2}} r^4+68
\\\nonumber
&\times&e^{\frac{A_{1} r^2}{R^2}} r^2-61\big) \eta ^2+12 \big(3
e^{\frac{2 A_{1} r^2}{R^2}} r^4+2 e^{\frac{A_{1} r^2}{R^2}}
r^2-4\big) \eta -12\big)-R^2 \big(4 e^{\frac{2 A_{1} r^2}{R^2}} \eta
\\\nonumber
&\times&(4 \eta +3)^2 r^4-4 e^{\frac{A_{1} r^2}{R^2}} \big(32 \eta
^3+84 \eta ^2+69 \eta +18\big) r^2-24 \eta ^3-61 \eta ^2-48 \eta
\\\nonumber
&-&12\big)\big) r^4+a \big(4 e^{\frac{A_{1} r^2}{R^2}} \big(11 \eta
^2+17 \eta +6\big) \big(e^{\frac{A_{1} r^2}{R^2}} \big(2 \eta ^2+6
\eta +3\big)-3 \big(2 \eta ^2\big)\big)
\\\nonumber
&\times& R^4-A_{1} \big(-4 e^{\frac{A_{1} r^2}{R^2}} \eta \big(148
\eta ^3+403 \eta ^2+339 \eta +90\big) r^2+4 e^{\frac{2 A_{1}
r^2}{R^2}} \big(2 \big(70 r^2
\\\nonumber
&-&11\big) \eta ^4+5 \big(49 r^2-20\big) \eta ^3+3 \big(43
r^2-49\big) \eta ^2+3 \big(6 r^2-29\big) \eta -18\big) r^2
\\\nonumber
&+&3 \big(40 \eta ^4+179 \eta ^3+273 \eta ^2+168 \eta +36\big)\big)
R^2+2 A_{1}^2 r^2 \eta  \big(8 \big(8 e^{\frac{2 A_{1} r^2}{R^2}}
r^4
\\\nonumber
&+&6 e^{\frac{A_{1} r^2}{R^2}} r^2-3\big) \eta ^3+\big(96 e^{\frac{2
A_{1} r^2}{R^2}} r^4+68 e^{\frac{A_{1} r^2}{R^2}} r^2-61\big) \eta
^2+12 \big(3 e^{\frac{2 A_{1} r^2}{R^2}} r^4
\\\nonumber
&+&2 e^{\frac{A_{1} r^2}{R^2}} r^2-4\big) \eta -12\big)\big) r^4+R^2
\big(-A_{1}^2 \big(104 \eta ^4+363 \eta ^3+441 \eta ^2-4
e^{\frac{A_{1} r^2}{R^2}} r^2
\\\nonumber
&\times& \big(52 \eta ^3+123 \eta ^2+87 \eta +18\big) \eta +216 \eta
+4 e^{\frac{2 A_{1} r^2}{R^2}} r^2 \big(\big(4 r^2-22\big) \eta
^4+\big(23 r^2
\\\nonumber
&-&100\big) \eta ^3+3 \big(13 r^2-49\big) \eta ^2+3 \big(6
r^2-29\big) \eta -18\big)+36\big) r^4-A_{1} R^2 \big(104 \eta ^4
\\\nonumber
&+&331 \eta ^3+381 \eta ^2+192 \eta -4 e^{\frac{A_{1} r^2}{R^2}} r^2
\big(74 \eta ^4+141 \eta ^3+54 \eta ^2-33 \eta -18\big)
\\\nonumber
&+&4 e^{\frac{2 A_{1} r^2}{R^2}} r^2 \big(4 r^2 \eta ^4+3 \big(5
r^2-22\big) \eta ^3+9 \big(r^2-15\big) \eta ^2-87 \eta
-18\big)+36\big) r^2
\\\nonumber
&+&\big(-1+e^{\frac{A_{1} r^2}{R^2}}\big) R^4 \big(8 \big(2
e^{\frac{2 A_{1} r^2}{R^2}} r^4-26 e^{\frac{A_{1} r^2}{R^2}}
r^2+13\big) \eta ^4+\big(60 e^{\frac{2 A_{1} r^2}{R^2}} r^4-428
\\\label{38}
&\times&e^{\frac{A_{1} r^2}{R^2}} r^2+\big) ^3+\big( e^{\frac{2
r^2}{R^2}} r^4-e^{\frac{A_{1} r^2}{R^2}} r^2+\big) \eta ^2-
\big(e^{\frac{A_{1} r^2}{R^2}} r^2-8\big) \eta
+\big)\big)\bigg]^{-1},
\end{eqnarray}
\begin{eqnarray}\nonumber
v_{t}^{2}&=&\bigg[(\eta +1) \big(A_{1}^2 \big(2 a \eta  \big(8
\big(2 e^{\frac{A_{1} r^2}{R^2}} r^2-1\big) \eta ^2+3 \big(4
e^{\frac{A_{1} r^2}{R^2}} r^2-5\big) \eta -6\big) r^2+R^2 \big(8
\\\nonumber
&\times&e^{\frac{A_{1} r^2}{R^2}} r^2 +e^{\frac{2 A_{1} r^2}{R^2}}
\big(10 r^4+r^2\big)+9\big) \eta ^3+3 \big(-100 e^{\frac{A_{1}
r^2}{R^2}} r^2+4 e^{\frac{2 A_{1} r^2}{R^2}} \big(5 r^2+2\big) r^2
\\\nonumber
&+&6 \big(-24 e^{\frac{A_{1} r^2}{R^2}} r^2+2 e^{\frac{2 A_{1}
r^2}{R^2}} r^2+39\big) \eta +72\big)\big) r^4+A_{1} \big(-2 a^2 \eta
\big(8 \big(2 e^{\frac{A_{1} r^2}{R^2}} r^2-1\big) \eta ^2
\\\nonumber
&+&3 \big(4 e^{\frac{A_{1} r^2}{R^2}} r^2-5\big) \eta -6\big) r^4-a
R^2 \eta \big(4 e^{\frac{A_{1} r^2}{R^2}} \eta (4 \eta +3) r^2+4
e^{\frac{2 A_{1} r^2}{R^2}} \big(\big(4 r^2-2\big)
\\\nonumber
&\times& \eta ^2+3 \big(r^2-2\big) \eta -3\big) r^2-3 \big(8 \eta
^2+15 \eta +6\big)\big) r^2+R^4 \big(-8 \big(6 e^{\frac{2 A_{1}
r^2}{R^2}} r^4
\\\nonumber
&-&15 e^{\frac{A_{1} r^2}{R^2}} r^2+7\big) \eta ^3-3 \big(-60
e^{\frac{A_{1} r^2}{R^2}} r^2+4 e^{\frac{2 A_{1} r^2}{R^2}} \big(3
r^2-2\big) r^2+51\big) \eta ^2+12
\\\nonumber
&\times& \big(5 e^{\frac{A_{1} r^2}{R^2}} r^2+e^{\frac{2 A_{1}
r^2}{R^2}} r^2-11\big) \eta -36\big)\big) r^2+R^2 \big(-2 a^2 \eta
\big(8 \eta ^2+15 \eta +6\big) r^4
\\\nonumber
&+&4 a e^{\frac{A_{1} r^2}{R^2}} R^2 \eta \big(e^{\frac{A_{1}
r^2}{R^2}} \big(2 \eta ^2+6 \eta +3\big)-3 \big(2 \eta ^2+3 \eta
+1\big)\big) r^4+\big(-1+e^{\frac{A_{1} r^2}{R^2}}\big)
\\\nonumber
&\times& R^4 \big(8 \big(6 e^{\frac{2 A_{1} r^2}{R^2}} r^4-14
e^{\frac{A_{1} r^2}{R^2}} r^2+7\big) \eta ^3+9 \big(4 e^{\frac{2
A_{1} r^2}{R^2}} r^4-20 e^{\frac{ A_{1}r^2}{R^2}} r^2+17\big) \eta
^2
\\\nonumber
&+&\big(132-72 e^{\frac{A_{1} r^2}{R^2}} r^2\big) \eta
+36\big)\big)\big)\bigg]\bigg[a^2 \big(A_{1} r^2 \big(16 \big(4
e^{\frac{2 A_{1} r^2}{R^2}} r^4-6 e^{\frac{A_{1} r^2}{R^2}}
r^2+3\big) \eta ^4
\\\nonumber
&+&2 \big(48 e^{\frac{2 A_{1} r^2}{R^2}} r^4-140 e^{\frac{A_{1}
r^2}{R^2}} r^2+97\big) \eta ^3+9 \big(4 e^{\frac{2 A_{1} r^2}{R^2}}
r^4-28 e^{\frac{A_{1} r^2}{R^2}} r^2+31\big) \eta ^2
\\\nonumber
&-&24 \big(3 e^{\frac{A_{1} r^2}{R^2}} r^2-7\big) \eta +36\big)-R^2
\big(16 \big(4 e^{\frac{2 A_{1} r^2}{R^2}} r^4-8 e^{\frac{A_{1}
r^2}{R^2}} r^2+3\big) \eta ^4+2
\\\nonumber
&\times& \big(48 e^{\frac{2 A_{1} r^2}{R^2}} r^4-168 e^{\frac{A_{1}
r^2}{R^2}} r^2+97\big) \eta ^3+3 \big(12 e^{\frac{2 f r^2}{R^2}}
r^4-92 e^{\frac{A_{1} r^2}{R^2}} r^2+93\big) \eta ^2
\\\nonumber
&-&24 \big(3 e^{\frac{A_{1} r^2}{R^2}} r^2-7\big) \eta +36\big)\big)
r^4-a \big(4 e^{\frac{A_{1} r^2}{R^2}} \eta (\eta +1)
\big(e^{\frac{A_{1} r^2}{R^2}} \big(2 \eta ^2+6 \eta +3\big)
\\\nonumber
&-&\big(2 \eta ^2+3 \eta +1\big)\big) R^4-A_{1} \big(8 \big(-46
e^{\frac{A_{1} r^2}{R^2}} r^2+e^{\frac{2 A_{1} r^2}{R^2}} \big(26
r^2-1\big) r^2+21\big) \eta ^4
\\\nonumber
&+&\big(-980 e^{\frac{A_{1} r^2}{R^2}} r^2+4 e^{\frac{2 A_{1}
r^2}{R^2}} \big(79 r^2-8\big) r^2+651\big) \eta ^3+12 \big(-68
e^{\frac{A_{1} r^2}{R^2}} r^2
\\\nonumber
&+&e^{\frac{2 A_{1} r^2}{R^2}} \big(10 r^2-3\big) r^2+75\big) \eta
^2-6 \big(36 e^{\frac{A_{1} r^2}{R^2}} r^2+2 e^{\frac{2 A_{1}
r^2}{R^2}} r^2-87\big) \eta +108\big) R^2
\\\nonumber
&+&A_{1}^2 r^2 \big(16 \big(4 e^{\frac{2 A_{1} r^2}{R^2}} r^4-6
e^{\frac{A_{1} r^2}{R^2}} r^2+3\big) \eta ^4+2 \big(48 e^{\frac{2
A_{1} r^2}{R^2}} r^4-140 e^{\frac{A_{1} r^2}{R^2}} r^2+97\big) \eta
^3
\\\nonumber
&+&9 \big(4 e^{\frac{2 A_{1} r^2}{R^2}} r^4-28 e^{\frac{A_{1}
r^2}{R^2}} r^2+31\big) \eta ^2-24 \big(3 e^{\frac{A_{1} r^2}{R^2}}
r^2-7\big) \eta +36\big)\big) r^4
\\\nonumber
&+&R^2 \big(-A_{1}^2 \big(8 \big(-2 e^{\frac{A_{1} r^2}{R^2}}
r^2+e^{\frac{2 A_{1} r^2}{R^2}} \big(2 r^4+r^2\big)+1\big) \eta
^4+\big(-108 e^{\frac{A_{1}r^2}{R^2}} r^2+4
\\\nonumber
&\times& e^{\frac{2 A_{1}r^2}{R^2}} \big(11 r^2+8\big) r^2+63\big)
\eta ^3+12 \big(-14 e^{\frac{A_{1} r^2}{R^2}} r^2+e^{\frac{2 A_{1}
r^2}{R^2}} \big(2 r^2+3\big) r^2+12\big) \eta ^2
\\\nonumber
&+&6 \big(-12 e^{\frac{A_{1} r^2}{R^2}} r^2+2 e^{\frac{2 A_{1}
r^2}{R^2}} r^2+21\big) \eta +36\big) r^4-A_{1} R^2 \eta \big(8 \eta
^3+31 \eta ^2+36 \eta -4
\\\nonumber
&\times& e^{\frac{A_{1} r^2}{R^2}} r^2 \big(2 \eta ^3+9 \eta ^2+9
\eta +3\big)+4 e^{\frac{2 A_{1} r^2}{R^2}} r^2 \big(4 r^2 \eta ^3+3
\big(r^2+2\big) \eta ^2+9 \eta +3\big)
\\\nonumber
&+&12\big) r^2+\big(-1+e^{\frac{A_{1} r^2}{R^2}}\big) R^4 \eta
\big(8 \big(2 e^{\frac{2 A_{1} r^2}{R^2}} r^4-2 e^{\frac{A_{1}
r^2}{R^2}} r^2+1\big) \eta ^3+\big(12 e^{\frac{2 A_{1} r^2}{R^2}}
r^4
\\\label{39}
&-&44 e^{\frac{A_{1} r^2}{R^2}} r^2+31\big) \eta ^2+\big(36-24
e^{\frac{A_{1} r^2}{R^2}} r^2\big) \eta +12\big)\big)\bigg]^{-1}.
\end{eqnarray}

\section*{Appendix B: Adiabatic Index}
\renewcommand{\theequation}{B\arabic{equation}}
\setcounter{equation}{0} The expressions of the adiabatic index is
given as follows.

\begin{eqnarray}\nonumber
\Gamma&=&\frac{4}{3} \bigg[1-\bigg\{3 R^2 \big(8 \big(e^{\frac{A_{1}
r^2}{R^2}} r^2-1\big) \eta ^2+3 \big(2 e^{\frac{A_{1} r^2}{R^2}}
r^2-5\big) \eta -6\big) \big(-a^2 \big(8 e^{\frac{A_{1} r^2}{R^2}}
r^2 \eta ^2
\\\nonumber
&+&\big(6 e^{\frac{A_{1} r^2}{R^2}} r^2-3\big) \eta -2\big) r^4+a
\big(A_{1} \big(8 e^{\frac{A_{1} r^2}{R^2}} r^2 \eta ^2+\big(6
e^{\frac{A_{1} r^2}{R^2}} r^2-3\big) \eta -2\big) r^2+2 R^2
\\\nonumber
&\times&\big(2 e^{\frac{2 A_{1} r^2}{R^2}} \big(2 \eta ^2+3 \eta
+1\big) r^2+6 \eta ^2+12 \eta -e^{\frac{A_{1} r^2}{R^2}} \big(2
\big(6 \eta ^2+6 \eta +1\big) r^2+2 \eta ^2
\\\nonumber
&+&3 \eta +1\big)+5\big)\big) r^2+R^2 \big(4 a \eta ^2 r^2+5 A_{1} \eta r^2+2
e^{\frac{2 A_{1} r^2}{R^2}} \big(2 A_{1} \big(2 \eta ^2+3 \eta +1\big) r^2
\\\nonumber
&+&R^2 \eta \big) r^2+2 R^2+4 R^2 \eta ^2+5 R^2 \eta -e^{\frac{A_{1}
r^2}{R^2}} \big(2 A_{1} \big((\eta +2) r^2+2 \eta ^2+3 \eta +1\big)
r^2
\\\nonumber
&+&R^2 \big(4 \eta ^2+\big(2 r^2+5\big) \eta +2\big)\big)\big)\big)\bigg\}\bigg\{2
\big(2 a^2 \eta \big(a r^2 \big(8 \big(8 e^{\frac{2 A_{1} r^2}{R^2}} r^4+6
e^{\frac{A_{1} r^2}{R^2}} r^2-3\big) \eta ^3
\\\nonumber
&+&\big(96 e^{\frac{2 A_{1} r^2}{R^2}} r^4+68 e^{\frac{A_{1}
r^2}{R^2}} r^2-61\big) \eta ^2+12 \big(3 e^{\frac{2 A_{1} r^2}{R^2}}
r^4+2 e^{\frac{A_{1} r^2}{R^2}} r^2-4\big) \eta -12\big)
\\\nonumber
&-&R^2 \big(4 e^{\frac{2 A_{1} r^2}{R^2}} \eta  (4 \eta +3)^2 r^4-4
e^{\frac{A_{1} r^2}{R^2}} \big(32 \eta ^3+84 \eta ^2+69 \eta
+18\big) r^2-24 \eta ^3
\\\nonumber
&-&61 \eta ^2-48 \eta -12\big)\big) r^4-a \big(4 e^{\frac{A_{1}
r^2}{R^2}} \big(11 \eta ^2+17 \eta +6\big) \big(e^{\frac{A_{1}
r^2}{R^2}} \big(2 \eta ^2+6 \eta +3\big)
\\\nonumber
&-&3 \big(2 \eta ^2+3\eta +1\big)\big) R^4-A_{1} \big(-4
e^{\frac{A_{1} r^2}{R^2}} \eta \big(148 \eta ^3+403 \eta ^2+339 \eta
+90\big) r^2
\\\nonumber
&+&4 e^{\frac{2 A_{1} r^2}{R^2}} \big(2 \big(70 r^2-11\big) \eta
^4+5 \big(49 r^2-20\big) \eta ^3+3 \big(43 r^2-49\big) \eta ^2
\\\nonumber
&+&3 \big(6 r^2-29\big) \eta -18\big) r^2+3 \big(40 \eta ^4+179 \eta
^3+273 \eta ^2+168 \eta +36\big)\big) R^2
\\\nonumber
&+&2 A_{1}^2 r^2 \eta \big(8 \big(8 e^{\frac{2 A_{1} r^2}{R^2}}
r^4+6 e^{\frac{A_{1} r^2}{R^2}} r^2-3\big) \eta ^3+\big(96
e^{\frac{2 A_{1} r^2}{R^2}} r^4+68 e^{\frac{A_{1} r^2}{R^2}}
r^2-61\big) \eta ^2
\\\nonumber
&+&12 \big(3 e^{\frac{2 A_{1} r^2}{R^2}} r^4+2 e^{\frac{A_{1}
r^2}{R^2}} r^2-4\big) \eta -12\big)\big) r^4+R^2 \big(A_{1}^2
\big(104 \eta ^4+363 \eta ^3+441 \eta ^2
\\\nonumber
&-&4 e^{\frac{A_{1} r^2}{R^2}} r^2 \big(52 \eta ^3+123 \eta ^2+87
\eta +18\big) \eta +216 \eta +4 e^{\frac{2 A_{1} r^2}{R^2}} r^2
\big(\big(4 r^2-22\big) \eta ^4
\\\nonumber
&+&\big(23 r^2-100\big) \eta ^3+3 \big(13 r^2-49\big) \eta ^2+3
\big(6 r^2-29\big) \eta -18\big)+36\big) r^4
\\\nonumber
&+&a R^2 \big(104 \eta ^4+331 \eta ^3+381 \eta ^2+192 \eta -4 e^{\frac{A_{1}
r^2}{R^2}} r^2 \big(74 \eta ^4+141 \eta ^3+54 \eta ^2
\\\nonumber
&-&33 \eta -18\big)+4 e^{\frac{2 A_{1} r^2}{R^2}} r^2 \big(4 r^2
\eta ^4+3 \big(5 r^2-22\big) \eta ^3+9 \big(r^2-15\big) \eta ^2-87
\eta -18\big)
\\\nonumber
&+&36\big) r^2-\big(-1+e^{\frac{A_{1} r^2}{R^2}}\big) R^4 \big(8
\big(2 e^{\frac{2 A_{1} r^2}{R^2}} r^4-26 e^{\frac{A_{1} r^2}{R^2}}
r^2+13\big) \eta ^4+\big(60 e^{\frac{2 A_{1} r^2}{R^2}} r^4
\\\nonumber
&-&428 e^{\frac{A_{1} r^2}{R^2}} r^2+331\big) \eta ^3+\big(36
e^{\frac{2 A_{1} r^2}{R^2}} r^4-300 e^{\frac{A_{1} r^2}{R^2}}
r^2+381\big) \eta ^2
\\\nonumber
&-&24 \big(3 e^{\frac{A_{1} r^2}{R^2}} r^2-8\big) \eta
+36\big)\big)\big)\bigg\}^{-1}\bigg],
\end{eqnarray}
\begin{eqnarray}\nonumber
\Gamma_{r}&=&\bigg[4 (\eta +1) \big(\eta  \big(2 e^{\frac{A_{1}
r^2}{R^2}} (4 \eta +3) r^2-8 \eta -15\big)-6\big) \big(A_{1}^2
\big(2 a \eta \big(8 \big(2 e^{\frac{A_{1} r^2}{R^2}} r^2
\\\nonumber
&-&1\big) \eta ^2+3 \big(4 e^{\frac{A_{1} r^2}{R^2}} r^2-5\big) \eta
-6\big) r^2+R^2 \big(8 \big(-18 e^{\frac{A_{1} r^2}{R^2}}
r^2+e^{\frac{2 A_{1} r^2}{R^2}} \big(10 r^4+r^2\big)
\\\nonumber
&+&9\big) \eta ^3+3 \big(-100 e^{\frac{A_{1} r^2}{R^2}} r^2+4
e^{\frac{2 A_{1} r^2}{R^2}} \big(5 r^2+2\big) r^2+77\big) \eta ^2+6
\big(-24 e^{\frac{A_{1} r^2}{R^2}} r^2
\\\nonumber
&+&2 e^{\frac{2 A_{1} r^2}{R^2}} r^2+39\big) \eta +72\big)\big) r^4+a \big(-2 a^2
\eta \big(8 \big(2 e^{\frac{A_{1} r^2}{R^2}} r^2-1\big) \eta ^2+3 \big(4
e^{\frac{A_{1} r^2}{R^2}} r^2
\\\nonumber
&-&5\big) \eta -6\big) r^4-a R^2 \eta \big(4 e^{\frac{A_{1}
r^2}{R^2}} \eta (4 \eta +3) r^2+4 e^{\frac{2 A_{1} r^2}{R^2}}
\big(\big(4 r^2-2\big) \eta ^2+3 \big(r^2
\\\nonumber
&-&2\big) \eta -3\big) r^2-3 \big(8 \eta ^2+15 \eta +6\big)\big)
r^2+R^4 \big(-8 \big(6 e^{\frac{2 A_{1} r^2}{R^2}} r^4-15
e^{\frac{A_{1} r^2}{R^2}} r^2
\\\nonumber
&+&7\big) \eta ^3-3 \big(-60 e^{\frac{A_{1} r^2}{R^2}} r^2+4
e^{\frac{2 A_{1} r^2}{R^2}} \big(3 r^2-2\big) r^2+51\big) \eta ^2+12
\big(5 e^{\frac{A_{1} r^2}{R^2}} r^2
\\\nonumber
&+&e^{\frac{2 A_{1} r^2}{R^2}} r^2-11\big) \eta -36\big)\big)
r^2+R^2 \big(-2 a^2 \eta \big(8 \eta ^2+15 \eta +6\big) r^4+4 a
e^{\frac{f r^2}{R^2}}
\\\nonumber
&\times& R^2 \eta \big(e^{\frac{A_{1} r^2}{R^2}} \big(2 \eta ^2+6
\eta +3\big )-3 \big(2 \eta ^2+3 \eta +1\big)\big)
r^4+\big(-1+e^{\frac{A_{1} r^2}{R^2}}\big)
\\\nonumber
&\times& R^4 \big(8 \big(6 e^{\frac{2 A_{1} r^2}{R^2}} r^4-14
e^{\frac{A_{1} r^2}{R^2}} r^2+7\big) \eta ^3+9 \big(4 e^{\frac{2
A_{1} r^2}{R^2}} r^4-20 e^{\frac{A_{1} r^2}{R^2}} r^2+17\big) \eta
^2
\\\nonumber
&+&\big(132-72 e^{\frac{A_{1} r^2}{R^2}} r^2\big) \eta
+36\big)\big)\big) \big(-a^2 \big(2 e^{\frac{A_{1} r^2}{R^2}}
r^2+1\big) \eta  (4 \eta +3) r^4
\\\nonumber
&+&a \big(A_{1} \big(2 e^{\frac{A_{1} r^2}{R^2}} r^2+1\big) \eta  (4
\eta +3) r^2+R^2 \big(6 e^{\frac{2 A_{1} r^2}{R^2}} \big(2 \eta ^2+3
\eta +1\big) r^2
\\\nonumber
&+&10 \eta ^2+21 \eta -e^{\frac{A_{1} r^2}{R^2}} \big(\big(28 \eta
^2+30 \eta +6\big) r^2+6 \eta ^2+9 \eta +3\big)+9\big)\big) r^2
\\\nonumber
&+&R^2 \big(a \big(6 e^{\frac{2 A_{1} r^2}{R^2}} \big(2 \eta ^2+3 \eta +1\big) r^2+2
\eta ^2+e^{\frac{A_{1} r^2}{R^2}} \big(r^2 \big(4 \eta ^2-6\big)-3 \big(2 \eta ^2
\\\nonumber
&+&3 \eta +1\big)\big)-3\big) r^2+\big(-1+e^{\frac{A_{1}
r^2}{R^2}}\big ) R^2 \big(2 \big(2 e^{\frac{A_{1} r^2}{R^2}}
r^2-5\big) \eta ^2+3 \big(2 e^{\frac{A_{1} r^2}{R^2}} r^2
\\\nonumber
&-&5\big) \eta -6\big)\big)\big)\bigg]\bigg[R^4 \big(8
\big(e^{\frac{A_{1} r^2}{R^2}} r^2-1\big) \eta ^2+3 \big(2
e^{\frac{A_{1} r^2}{R^2}} r^2-5\big) \eta -6\big) \big(-2 a^2 \eta
\\\nonumber
&\times& \big(A_{1} r^2 \big(8 \big(8 e^{\frac{2 A_{1} r^2}{R^2}}
r^4+6 e^{\frac{A_{1} r^2}{R^2}} r^2-3\big) \eta ^3+\big(96
e^{\frac{2 A_{1} r^2}{R^2}} r^4+68 e^{\frac{A_{1} r^2}{R^2}}
r^2-61\big ) \eta ^2
\\\nonumber
&+&12 \big(3 e^{\frac{2 A_{1} r^2}{R^2}} r^4+2 e^{\frac{A_{1}
r^2}{R^2}} r^2-4\big) \eta -12\big)-R^2 \big(4 e^{\frac{2
A_{1}r^2}{R^2}} \eta (4 \eta +3)^2 r^4-4 e^{\frac{A_{1} r^2}{R^2}}
\\\nonumber
&\times& \big(32 \eta ^3+84 \eta ^2+69 \eta +18\big) r^2-24 \eta
^3-61 \eta ^2-48 \eta -12\big)\big) r^4+a \big(4 e^{\frac{A_{1}
r^2}{R^2}}
\\\nonumber
&\times& \big(11 \eta ^2+17 \eta +6\big) \big(e^{\frac{A_{1}
r^2}{R^2}} \big(2 \eta ^2+6 \eta +3\big)-3 \big(2 \eta ^2+3 \eta
+1\big )\big) R^4-A_{1}
\\\nonumber
&\times& \big(-4 e^{\frac{A_{1} r^2}{R^2}} \eta \big(148 \eta ^3+403
\eta ^2+339 \eta +90\big) r^2+4 e^{\frac{2 A_{1} r^2}{R^2}} \big(2
\big(70 r^2-11\big) \eta ^4
\\\nonumber
&+&5 \big(49 r^2-20\big) \eta ^3+3 \big(43 r^2-49\big ) \eta ^2+3
\big(6 r^2-29\big) \eta -18\big) r^2+3 \big(40 \eta ^4
\\\nonumber
&+&179 \eta ^3+273 \eta ^2+168 \eta +36\big)\big) R^2+2 A_{1}^2 r^2
\eta \big(8 \big(8 e^{\frac{2 A_{1} r^2}{R^2}} r^4+6 e^{\frac{A_{1}
r^2}{R^2}} r^2-3\big) \eta ^3
\\\nonumber
&+&\big(96 e^{\frac{2 A_{1} r^2}{R^2}} r^4+68 e^{\frac{A_{1}
r^2}{R^2}} r^2-61\big) \eta ^2+12 \big(3 e^{\frac{2 A_{1} r^2}{R^2}}
r^4+2 e^{\frac{A_{1} r^2}{R^2}} r^2-4\big) \eta -12\big)\big) r^4
\\\nonumber
&+&R^2 \big(-A_{1}^2 \big(104 \eta ^4+363 \eta ^3+441 \eta ^2-4
e^{\frac{A_{1} r^2}{R^2}} r^2 \big(52 \eta ^3+123 \eta ^2+87 \eta
+18\big) \eta
\\\nonumber
&+&216 \eta +4 e^{\frac{2 A_{1} r^2}{R^2}} r^2 \big(\big(4
r^2-22\big) \eta ^4+\big(23 r^2-100\big) \eta ^3+3 \big(13
r^2-49\big) \eta ^2
\\\nonumber
&+&3 \big(6 r^2-29\big) \eta -18\big)+36\big) r^4-A_{1} R^2 \big(104
\eta ^4+331 \eta ^3+381 \eta ^2+192 \eta
\\\nonumber
&-&4 e^{\frac{f r^2}{R^2}} r^2 \big(74 \eta ^4+141 \eta ^3+54 \eta
^2-33 \eta -18\big)+4 e^{\frac{2 A_{1} r^2}{R^2}} r^2 \big(4 r^2
\eta ^4+3 \big(5 r^2
\\\nonumber
&-&22\big) \eta ^3+9 \big(r^2-15\big) \eta ^2-87 \eta
-18\big)+36\big) r^2+\big(-1+e^{\frac{A_{1} r^2}{R^2}}\big) R^4
\\\nonumber
&\times& \big(8 \big(2 e^{\frac{2 A_{1} r^2}{R^2}} r^4-26
e^{\frac{A_{1} r^2}{R^2}} r^2+13\big) \eta ^4+\big(60 e^{\frac{2
A_{1} r^2}{R^2}} r^4-428 e^{\frac{A_{1} r^2}{R^2}} r^2+331\big) \eta
^3
\\\nonumber
&+&\big(36 e^{\frac{2 A_{1} r^2}{R^2}} r^4-300 e^{\frac{A_{1}
r^2}{R^2}} r^2+381\big) \eta ^2-24 \big(3 e^{\frac{A_{1} r^2}{R^2}}
r^2-8\big) \eta +36\big)\big)\big)
\\\nonumber
&\times& \bigg\{2 \big(-1+e^{\frac{A_{1} r^2}{R^2}}\big) \big(\eta
\big(2 e^{\frac{A_{1} r^2}{R^2}} (\eta +3) r^2-13 \eta
-17\big)-6\big)+\frac{1}{R^4}\bigg\{2 r^2
\\\nonumber
&\times& \big(-6 a^2 \eta ^2 r^2+6 a A_{1} \eta ^2 r^2-4 a^2 \eta
r^2+4 a A_{1} \eta r^2+2 e^{\frac{2 A_{1} r^2}{R^2}} (a+A_{1}) R^2
(\eta +1)
\\\nonumber
&\times&(11 \eta +6)r^2+18 a R^2+6 A_{1} R^2+21 a R^2 \eta ^2+13
A_{1}R^2 \eta ^2+43 a R^2 \eta
\\\nonumber
&+&+21 A_{1} R^2 \eta-e^{\frac{A_{1} r^2}{R^2}} \big(4 a^2 \eta  (4
\eta +3) r^4+A_{1} R^2 \big(2 (\eta +2) (\eta +3) r^2
\\\nonumber
&+&(\eta +1)(11 \eta +6)\big)+a \big(R^2 \big(2 (\eta (27 \eta
+29)+6) r^2+(\eta +1) (11 \eta +6)\big)
\\\nonumber
&-&4 A_{1} r^4 \eta  (4 \eta
+3)\big)\big)\big)\bigg\}\bigg\}\bigg]^{-1},
\end{eqnarray}
\begin{eqnarray}\nonumber
\Gamma_{t}&=&-\bigg[2 (\eta +1) \big(A_{1}^2 \big(2 a \eta  \big(8
\big(2 e^{\frac{f r^2}{R^2}} r^2-1\big) \eta ^2+3 \big(4
e^{\frac{A_{1} r^2}{R^2}} r^2-5\big) \eta -6\big) r^2
\\\nonumber
&+&R^2 \big(8 \big(-18 e^{\frac{A_{1} r^2}{R^2}} r^2+e^{\frac{2
A_{1} r^2}{R^2}} \big(10 r^4+r^2\big)+9\big) \eta ^3+3 \big(-100
e^{\frac{A_{1} r^2}{R^2}} r^2
\\\nonumber
&+&4 e^{\frac{2 A_{1} r^2}{R^2}} \big(5 r^2+2\big) r^2+77\big) \eta
^2+6 \big(-24 e^{\frac{A_{1} r^2}{R^2}} r^2+2 e^{\frac{2
A_{1}r^2}{R^2}} r^2+39\big) \eta +72\big)\big) r^4
\\\nonumber
&+&A_{1} \big(-2 a^2 \eta \big(8 \big(2 e^{\frac{A_{1} r^2}{R^2}}
r^2-1\big) \eta ^2+3 \big(4 e^{\frac{A_{1} r^2}{R^2}} r^2-5\big)
\eta -6\big) r^4-a R^2 \eta \big(4 e^{\frac{A_{1} r^2}{R^2}}
\\\nonumber
&\times& \eta  (4 \eta +3) r^2+4 e^{\frac{2 A_{1} r^2}{R^2}}
\big(\big(4 r^2-2\big) \eta ^2+3 \big(r^2-2\big) \eta -3\big) r^2-3
\big(8 \eta ^2
\\\nonumber
&+&15 \eta +6\big)\big) r^2+R^4 \big(-8 \big(6 e^{\frac{2
A_{1}r^2}{R^2}} r^4-15 e^{\frac{A_{1} r^2}{R^2}} r^2+7\big) \eta
^3-3 \big(-60 e^{\frac{A_{1} r^2}{R^2}} r^2
\\\nonumber
&+&4 e^{\frac{2 A_{1} r^2}{R^2}} \big(3 r^2-2\big) r^2+51\big) \eta
^2+12 \big(5 e^{\frac{A_{1} r^2}{R^2}} r^2+e^{\frac{2 A_{1}
r^2}{R^2}} r^2-11\big) \eta -36\big)\big) r^2
\\\nonumber
&+&R^2 \big(-2 a^2 \eta \big(8 \eta ^2+15 \eta +6\big) r^4+4 a
e^{\frac{A_{1} r^2}{R^2}} R^2 \eta \big(e^{\frac{A_{1} r^2}{R^2}}
\big(2 \eta ^2+6 \eta +3\big)
\\\nonumber
&-&3 \big(2 \eta ^2+3 \eta +1\big)\big) r^4+\big(-1+e^{\frac{A_{1}
r^2}{R^2}}\big) R^4 \big(8 \big(6 e^{\frac{2 A_{1} r^2}{R^2}} r^4-14
e^{\frac{A_{1} r^2}{R^2}} r^2+7\big) \eta ^3
\\\nonumber
&+&9 \big(4 e^{\frac{2 A_{1} r^2}{R^2}} r^4-20 e^{\frac{A_{1}
r^2}{R^2}} r^2+17\big) \eta ^2+\big(132-72 e^{\frac{A_{1} r^2}{R^2}}
r^2\big) \eta +36\big)\big)\big) \big(-a^2
\\\nonumber
&\times& \big(2 e^{\frac{A_{1} r^2}{R^2}} r^2+1\big) \eta  (4 \eta
+3) r^4+a \big(A_{1} \big(2 e^{\frac{A_{1} r^2}{R^2}} r^2+1\big)
\eta (4 \eta +3) r^2+R^2 \big(6 e^{\frac{2 f r^2}{R^2}}
\\\nonumber
&\times& \big(2 \eta ^2+3 \eta +1\big) r^2+10 \eta ^2+21 \eta
-e^{\frac{A_{1} r^2}{R^2}} \big(\big(28 \eta ^2+30 \eta +6\big)
r^2+6 \eta ^2
\\\nonumber
&+&9 \eta +3\big)+9\big)\big) r^2+R^2 \big(A_{1} \big(6 e^{\frac{2
A_{1} r^2}{R^2}} \big(2 \eta ^2+3 \eta +1\big) r^2+2 \eta
^2+e^{\frac{A_{1} r^2}{R^2}} \big(r^2
\\\nonumber
&\times& \big(4 \eta ^2-6\big)-3 \big(2 \eta ^2+3 \eta
+1\big)\big)-3\big) r^2+\big(-1+e^{\frac{A_{1} r^2}{R^2}}\big) R^2
\big(2 \big(2 e^{\frac{A_{1} r^2}{R^2}} r^2
\\\nonumber
&-&5\big) \eta ^2+3 \big(2 e^{\frac{A_{1} r^2}{R^2}} r^2-5\big) \eta
-6\big)\big)\big)\bigg]\bigg[\big(a^2 \big(R^2 \big(16 \big(4
e^{\frac{2 A_{1} r^2}{R^2}} r^4-8 e^{\frac{A_{1} r^2}{R^2}}
r^2+3\big) \eta ^4
\\\nonumber
&+&2 \big(48 e^{\frac{2 A_{1} r^2}{R^2}} r^4-168 e^{\frac{A_{1}
r^2}{R^2}} r^2+97\big) \eta ^3+3 \big(12 e^{\frac{2 A_{1} r^2}{R^2}}
r^4-92 e^{\frac{A_{1} r^2}{R^2}} r^2+93\big) \eta ^2
\\\nonumber
&-&24 \big(3 e^{\frac{A_{1} r^2}{R^2}} r^2-7\big) \eta
+36\big)-A_{1} r^2 \big(16 \big(4 e^{\frac{2 A_{1} r^2}{R^2}} r^4-6
e^{\frac{A_{1} r^2}{R^2}} r^2+3\big) \eta ^4+2 \big(48
\\\nonumber
&\times& e^{\frac{2 A_{1} r^2}{R^2}} r^4-140 e^{\frac{A_{1}
r^2}{R^2}} r^2+97\big) \eta ^3+9 \big(4 e^{\frac{2 A_{1} r^2}{R^2}}
r^4-28 e^{\frac{A_{1} r^2}{R^2}} r^2+31\big) \eta ^2
\\\nonumber
&-&24 \big(3 e^{\frac{A_{1} r^2}{R^2}} r^2-7\big) \eta +36\big)\big)
r^4+a \big(4 e^{\frac{A_{1} r^2}{R^2}} \eta (\eta +1)
\big(e^{\frac{A_{1} r^2}{R^2}} \big(2 \eta ^2+6 \eta +3\big)
\\\nonumber
&-&3 \big(2 \eta ^2+3 \eta +1\big)\big) R^4-A_{1} \big(8 \big(-46
e^{\frac{A_{1} r^2}{R^2}} r^2+e^{\frac{2 A_{1} r^2}{R^2}} \big(26
r^2-1\big) r^2+21\big) \eta ^4
\\\nonumber
&\times&\big(-980 e^{\frac{A_{1} r^2}{R^2}} r^2+4 e^{\frac{2 A_{1}
r^2}{R^2}} \big(79 r^2-8\big) r^2+651\big) \eta ^3+12 \big(-68
e^{\frac{A_{1} r^2}{R^2}} r^2
\\\nonumber
&+&e^{\frac{2 A_{1} r^2}{R^2}} \big(10 r^2-3\big) r^2+75\big) \eta
^2-6 \big(36 e^{\frac{A_{1} r^2}{R^2}} r^2+2 e^{\frac{2 A_{1}
r^2}{R^2}} r^2-87\big) \eta +108\big) R^2
\\\nonumber
&+&a^2 r^2 \big(16 \big(4 e^{\frac{2 A_{1} r^2}{R^2}} r^4-6 e^{\frac{A_{1} r^2}{R^2}}
r^2+3\big) \eta ^4+2 \big(48 e^{\frac{2 A_{1} r^2}{R^2}} r^4-140 e^{\frac{A_{1}
r^2}{R^2}} r^2+97\big) \eta ^3
\\\nonumber
&+&9 \big(4 e^{\frac{2 A_{1} r^2}{R^2}} r^4-28 e^{\frac{A_{1}
r^2}{R^2}} r^2+31\big) \eta ^2-24 \big(3 e^{\frac{A_{1} r^2}{R^2}}
r^2-7\big) \eta +36\big)\big) r^4+R^2 \big(A_{1}^2
\\\nonumber
&\times& \big(8 \big(-2 e^{\frac{A_{1} r^2}{R^2}} r^2+e^{\frac{2
A_{1} r^2}{R^2}} \big(2 r^4+r^2\big)+1\big) \eta ^4+\big(-108
e^{\frac{A_{1} r^2}{R^2}} r^2+4 e^{\frac{2 A_{1} r^2}{R^2}}
\\\nonumber
&\times& \big(11 r^2+8\big) r^2+63\big) \eta ^3+12 \big(-14
e^{\frac{A_{1} r^2}{R^2}} r^2+e^{\frac{2 A_{1} r^2}{R^2}} \big(2
r^2+3\big) r^2+12\big) \eta ^2
\\\nonumber
&+&6 \big(-12 e^{\frac{A_{1} r^2}{R^2}} r^2+2 e^{\frac{2
A_{1}r^2}{R^2}} r^2+21\big) \eta +36\big) r^4+A_{1} R^2 \eta \big(8
\eta ^3+31 \eta ^2+36 \eta
\\\nonumber
&-&4 e^{\frac{A_{1} r^2}{R^2}} r^2 \big(2 \eta ^3+9 \eta ^2+9 \eta
+3\big)+4 e^{\frac{2 A_{1} r^2}{R^2}} r^2 \big(4 r^2 \eta ^3+3
\big(r^2+2\big) \eta ^2+9 \eta +3\big)
\\\nonumber
&+&12\big) r^2-\big(-1+e^{\frac{A_{1} r^2}{R^2}}\big) R^4 \eta \big(8 \big(2
e^{\frac{2 a r^2}{R^2}} r^4-2 e^{\frac{A_{1} r^2}{R^2}} r^2+1\big) \eta ^3+\big(12
e^{\frac{2 A_{1} r^2}{R^2}} r^4
\\\nonumber
&-&44 e^{\frac{A_{1} r^2}{R^2}} r^2+31\big) \eta ^2+\big(36-24
e^{\frac{A_{1} r^2}{R^2}} r^2\big) \eta +12\big)\big)\big) \big(-2
a^2 \eta \big(2 e^{\frac{A_{1} r^2}{R^2}} (4 \eta +3) r^2
\\\nonumber
&+&3 \eta +2\big) r^4+a \big(2 A_{1} \eta \big(2 e^{\frac{A_{1}
r^2}{R^2}} (4 \eta +3) r^2+3 \eta +2\big) r^2+R^2 \big(2 e^{\frac{2
f r^2}{R^2}} \big(11 \eta ^2
\\\nonumber
&+&17 \eta +6\big) r^2+21 \eta ^2+43 \eta -e^{\frac{A_{1} r^2}{R^2}}
\big(2 \big(27 \eta ^2+29 \eta +6\big) r^2+11 \eta ^2+17 \eta
+6\big)
\\\nonumber
&+&18\big)\big) r^2+R^2 \big(A_{1} \big(2 e^{\frac{2 A_{1}
r^2}{R^2}} \big(11 \eta ^2+17 \eta +6\big) r^2+13 \eta ^2+21 \eta
-e^{\frac{A_{1} r^2}{R^2}} \big(2 \big(\eta ^2
\\\nonumber
&+&5 \eta +6\big) r^2+11 \eta ^2+17 \eta +6\big)+6\big)
r^2+\big(-1+e^{\frac{A_{1} r^2}{R^2}}\big) R^2 \big(\big(2
e^{\frac{A_{1} r^2}{R^2}} r^2
\\\nonumber
&-&13\big) \eta ^2+\big(6 e^{\frac{A_{1} r^2}{R^2}} r^2-17\big) \eta
-6\big)\big)\big)\bigg].
\end{eqnarray}

\section*{Appendix C:\label{tab:symbols}}
\renewcommand{\theequation}{C\arabic{equation}}
\setcounter{equation}{0} The comprehensive table of symbols is given
as follows.
\begin{table}[h!]
\caption{Definitions and first appearances of symbols used in the
manuscript.} \label{tab:symbols} \vspace{0.5cm} \centering
\begin{tabular}{|c|p{4cm}|p{4cm}|c|}
\hline \textbf{Symbol} & \textbf{Meaning} & \textbf{Context} &
\textbf{First Appearance}
\\ \hline
$\mathcal{Q}$ & Non-metricity scalar & Measures deviation from
metricity & Abstract
\\ \hline
$\mathbb{T}$ & Trace of the energy-momentum tensor & Summarizes
matter contribution & Abstract
\\ \hline
$\mathcal{Q}_{\rho\xi\mu}$ &non-metricity tensor & Describes
geometric deformation & Eq.(3)
\\ \hline
$P_{\rho\xi\mu}$ & Superpotential related to non-metricity & Derived
from non-metricity contributions & Eq.(14)
\\ \hline
$f(\mathcal{Q}, \mathbb{T})$ & An arbitrary function of
$\mathcal{Q}$ and $\mathbb{T}$ & Governs the dynamics of the gravity
model & Abstract
\\ \hline
$L_{m}$ & Matter Lagrangian & Represents matter content & Eq.(9)
\\ \hline
$\sigma = P_t - P_r$ & Anisotropy factor & Quantifies the pressure
anisotropy & Eq.(21)
\\ \hline
$\zeta, \eta$ & Parameters in $f(\mathcal{Q}, \mathbb{T})$ &
Coupling constants linking gravity and matter & Eq.(27)
\\ \hline
$A_1, a_1, a$ & Krori-Barua metric parameters & Define the metric
potentials $\alpha(r)$, $\beta(r)$ & Eq.(31)
\\ \hline
$u$ & Compactness parameter & Relates mass to radius: $u =
\frac{2M}{R}$ & Eq.(36)
\\ \hline
$v^2_r, v^2_t$ & Radial and tangential sound speeds & Measure sound
propagation speeds & Eq.(42)\\ \hline
\end{tabular}
\end{table}
\\
\textbf{Data availability:} No new data were generated or analyzed
in support of this research.

\end{document}